\documentclass[letterpaper]{article}
\usepackage{proceed2e}

\usepackage[margin=0.68in]{geometry}

\usepackage[usenames,dvipsnames]{xcolor}
\usepackage[graphicx]{realboxes}
\usepackage{rotating}
\usepackage{booktabs}
\usepackage{bbding}
\usepackage{subfigure}
\usepackage{tabularx}
\usepackage{multirow}
\usepackage{array}
\usepackage{upgreek}
\usepackage{bigints}
\usepackage{epsfig}
\usepackage{mdwlist}
\usepackage{flushend}
\usepackage{array}
\usepackage{bm}
\usepackage{amssymb}
\usepackage{amsmath}
\usepackage{amsfonts}
\usepackage{mathrsfs}
\usepackage{pifont}
\usepackage{lipsum}

\usepackage[bookmarks=true,
			colorlinks=true,
			linkcolor=RoyalBlue, 
			urlcolor=RoyalBlue,  
			citecolor=RoyalBlue]{hyperref}
\usepackage{url}

\usepackage{soul}

\usepackage{algorithm}
\usepackage{algpseudocode}

\usepackage{natbib}
\setcitestyle{authoryear}

\usepackage{times}

\usepackage{makecell}
\newcolumntype{x}[1]{>{\centering\arraybackslash}p{#1}}
\usepackage{tikz}

\usepackage[labelfont=bf,font=bf]{caption}

\title{Combining guaranteed and spot markets in display advertising: Selling guaranteed page views with stochastic demand\thanks{~\hspace*{5pt} B.Chen, J.Huang, Y.Huang, S.Kollias, and S.Yue. Combining guaranteed and spot markets in display advertising: selling guaranteed page views with stochastic demand. \emph{European Journal of Operational Research}, 280(3):1144–1159, 2020. [\href{https://www.sciencedirect.com/science/article/abs/pii/S0377221719306460}{Publication Link}]}}

\author{ 
{\bf Bowei Chen\thanks{~\hspace*{5pt}Corresponding author}}\\
Adam Smith Business School\\ 
University of Glasgow\\
\href{mailto:bowei.chen@glasgow.ac.uk}{bowei.chen@glasgow.ac.uk}\\
\And
{\bf Jingmin Huang}  \\
School of Computing Science      \\
University of Glasgow \\
\href{mailto:jingmin.huang@glasgow.ac.uk}{jingmin.huang@glasgow.ac.uk}\\
\And
{\bf Yufei Huang}  \\
Trinity Business School      \\
Trinity College Dublin \\
\href{mailto:yufei.huang@tcd.ie}{yufei.huang@tcd.ie}\\
\And
{\bf Stefanos Kollias, Shigang Yue}  \\
School of Computer Science      \\
University of Lincoln \\
\href{mailto:skollias@lincoln.ac.uk;syue@lincoln.ac.uk}{\{skollias,syue\}@lincoln.ac.uk}\\
}

\begin{document}

\maketitle

\begin{abstract}
While page views are often sold instantly through real-time auctions when users visit websites, they can also be sold in advance via guaranteed contracts. In this paper, we present a dynamic programming model to study how an online publisher should optimally allocate and price page views between guaranteed and spot markets. The problem is challenging because the allocation and pricing of guaranteed contracts affect how advertisers split their purchases between the two markets, and the terminal value of the model is endogenously determined by the updated dual force of supply and demand in auctions. We take the advertisers’ purchasing behaviour into consideration, i.e., risk aversion and stochastic demand arrivals, and present a scalable and efficient algorithm for the optimal solution. The model is also empirically validated with a commercial dataset. The experimental results show that selling page views via both channels can increase the publisher’s expected total revenue, and the optimal pricing and allocation strategies are robust to different market and advertiser types.
\end{abstract}


\section{Introduction}

Display advertising is one of the most popular forms of online marketing. It uses the Internet and the World Wide Web as an advertising medium, and when users visit websites, the promotional messages (i.e., the \emph{ads}), appear on the pages. They usually come in terms of rectangular images or photos placed on a web page either above, below or on the sides of the page's main content and are linked to other web pages. Online publishers make profits by selling the page views, namely the \emph{impressions}, through two channels: (i) selling them in advance via contracts; or (ii) auctioning them off in real time when users visit the web pages. The former is called \textit{guaranteed contracts} (or \emph{reservation contracts}) while the latter is called \emph{real-time bidding} (RTB). Over the past decades, RTB has become the widely used sales model for display advertising, in which advertisers come to a common marketplace, i.e., ad exchange, to compete for impressions from their targeted users~\citep{Muthukrishnan_2009,Mansour_2012}. It is real-time, impression-level and auction-based, thus has achieved a significant level of automation, integration and user-targeting~\citep{YYuan_2014,Sun_2016}.

Although RTB is more widely used, guaranteed contracts in fact have a longer history. In 1994, wire.com signed fourteen contracts with companies, such as AT\&T, Club Med and Coor'z Zima, being recognised as the start of display advertising~\citep{DoubleClick_2005}. A guaranteed contract is an agreement and it is usually negotiated privately between a publisher and an advertiser for bulk sales. Only a small portion of impressions on the market is sold through guaranteed contracts but they bring in much more revenue than RTB~\citep{eMarketer_2013_RTB}. In order to meet the demand for automation due to the huge number of site visits, standardised guaranteed contracts have been recently discussed. This is known as \emph{programmatic guarantee} (PG). In essence, PG is a sales system that sells future impressions via standardised guaranteed contracts in addition to RTB~\citep{OpenX_2013}. Examples include Google DoubleClick's Programmatic Guaranteed, AOL's Programmatic Upfront and Rubicon Project's Reserved Premium Media Buys. Recent studies have investigated PG from different perspectives and we provide an extensive review in Section~\ref{sec:related_work}. These studies aim to answer the following two main questions: (i) how many future impressions should be allocated to guaranteed contracts? and (ii) how to price the guaranteed contracts?

In this paper, we use a revenue maximisation model to study how to sell impressions using RTB and PG. More specifically, some impressions are sold in advance via standardised guaranteed contracts before the delivery day while the rest of the impressions are auctioned off in RTB using the second-price auctions~\citep{Ben-Zwi_2015}. We focus on one ad slot and study how the estimated total impressions in a future period, i.e., the ad delivery day, should be allocated and priced algorithmically between guaranteed and spot markets. For example, AOL sells the impressions from the top banner of its homepage on Christmas Day. AOL needs to decide on how many impressions should be sold a couple of days, weeks or months before Christmas via guaranteed contracts and at what prices. Unlike the traditional way of selling guaranteed contracts, in our model, there is no negotiation process between the publisher and the advertiser. Instead, the guaranteed contract price is posted in a common marketplace. Advertisers can monitor the price trend over time and purchase the needed impressions directly at the corresponding prices prior to the delivery day. Based on auction theory and operations research studies, we also consider the distinct characteristics of advertisers, i.e., risk aversion and stochastic arrivals, and then propose an algorithm to find the optimal allocation and pricing strategy based on the solution framework to the Knapsack problem. Our solution is a greedy algorithm and there are two significant differences to the Knapsack problem solution. First, for a given allocated purchase demand at a time point, the corresponding guaranteed contract price is obtained by considering advertiser's purchase behaviour. Second, the terminal value of RTB is not certain, which is determined by the updated dual force of supply and demand. The proposed model is further examined with an RTB dataset from a UK supply-side platform (SSP).\footnote{Supply-side platforms (SSPs) are intermediaries who help publishers sell impressions in RTB. Through SSPs, publishers are able to connect with advertisers from various ad exchanges and networks.} Our results show that introducing guaranteed contacts in addition to RTB increases the publisher's expected total revenue. Specifically, for ad slots with a high competition level, the guaranteed contract price significantly increases over time and the publisher allocates a large percentage of future impressions into guaranteed contracts, consequently, the revenue is mainly collected by PG. For ad slots with a low competition level, the guaranteed contract price increases steadily and less contracts are sold. However, our results show that the publisher's revenue from those ad slots can be significantly increased because there is a greater margin to be optimised. We further examine the robustness of our model by considering two extensions: (i) incorporating uncertainty into demand and supply of the impressions; and (ii) segmenting advertisers based on their valuations. Our analysis shows the model is robust under different conditions and the publisher can always obtain a higher total revenue when selling through both channels.

Our research makes the following contributions. First, different from the existing work which focuses on either guaranteed contracts or RTB, this paper is among the first to introduce a unified framework that combines PG and RTB simultaneously in display advertising. Second, we also study the interaction of pricing and allocation decisions on guaranteed contracts, thus can provide further insights to the existing literature that only focuses on either pricing or allocation. Third, different from the widely used conventional methodologies in marketing and operational studies, this paper proposes a data-driven analytical model. Our solution to the revenue maximisation problem is simple, efficient and scalable. The insights from our model are further validated using a commercial dataset. The robustness of the results indicates the potential usefulness of our model in practice.

The rest of the paper is structured as follows. Section~\ref{sec:related_work} reviews the related literature. Section~\ref{sec:model} discusses the model, including problem formulation, model assumptions and our solution. Section~\ref{sec:data} describes the used dataset and experimental settings. Section~\ref{sec:empirical_results} presents our experimental results and Section~\ref{sec:conclusion} concludes the paper.

\section{Related work}
\label{sec:related_work}

Our paper focuses on selling impressions via both RTB and PG, thus it naturally lies in the interface between mechanism design for online advertising auction and revenue management with dynamic pricing.

Mechanism design for online advertising has been extensively studied in the literature. Many discussions have been centred around search advertising auctions, such as the generalized first-price auction~\citep{Edelman_2007_2}, the generalized second-price auction~\citep{Edelman_2007_2,Lahaie_2011,Lahaie_2007,Varian_2007}, the Vickrey-Clarke-Groves (VCG) auction~\citep{Parkes_2007,Varian_2009,Varian_2014} and the optimal auction~\citep{Feldman_2010,Ostrovsky_2011,Thompson_2013} which extends Myerson's optimal auction for a single indivisible good~\citep{Myerson_1981}. Most of these studies on advertising auctions examine the properties of an auction model with respect to incentive compatibility, expected revenue, individual rationality, and computational complexity. In this paper, we focus on display advertising with the second-price auction. It has a simplified scenario in auction mechanism design because: (i) the measurement model is based on ad display rather than click, so click-through rate is not a major factor; (ii) ad slots on the same web page for a single page view is auctioned off separately so that each independent auction is a single-item auction.

Our paper is also related to the literature in revenue management~\citep{Bitran_2003,Talluri_2004}, in which many studies focus on how a seller uses dynamic pricing models to produce or offer a menu of products or services to its customers. For example, \cite{Gallego_1994} used intensity control to sell a given stock of products by a deadline when demand is price sensitive and stochastic and the seller\rq{}s objective is to maximise his expected revenue. Their model fits many applications such as single-route flight tickets selling and hotel rooms booking. \cite{Anjos_2004,Anjos_2005} proposed a dynamic pricing framework for selling flight tickets under the assumption of static demand. Our problem setting for PG is similar to the existing literature, however, the terminal value in our case is uncertain because the remaining impressions are auctioned off in RTB.

Existing literature has also studied selling products or services via both auctions and posted prices. For example, \cite{Caldentey_2007} discussed a problem of two channels selling, where the products can be sold through either an auction or an alternative channel with a posted price. They considered two scenarios of this dual-channel optimisation problem: in the first scenario, the posted price is an external channel run by another company; in the second scenario, the seller manages both auction and posted price channels. The second scenario is similar to our model setting. However, their discussion is mainly about the static posted price and they assume that the original values are uniformly distributed and there is no penalty cost. \cite{Gao_2016} studied a hybrid model that unifies both future and spot markets for dynamic spectrum access, in which buyers can purchase under-utilized licensed spectrum either through predefined contracts or through spot transactions with a VCG-like auction model. Their work is similar to ours, however, the seller does not optimise the contract price dynamically.

There are also a number of studies discussing the dual-channel problem in the context of display advertising. \cite{Feldman_2009} proposed a selection and matching algorithm for display ads, in their paper, the publisher's objective is not only to fulfil guaranteed contracts but also to deliver well-targeted impressions to advertisers. \cite{Ghosh_2009} proposed that a publisher can act as a bidder to bid for guaranteed contracts -- the allocation of impressions becomes the competition between the publisher and other advertisers so that impressions can be allocated to auctions only if advertisers' bids are high enough. \cite{Roels_2009}, \cite{Salomatin_2012}, \cite{Balseiro_2014} and \cite{Chen_2017_HKUST} discussed the optimal allocation using stochastic control models. \cite{Bharadwaj_2012} proposed a lightweight allocation framework which lets the real servers to allocate ads efficiently and with little overhead. \cite{Chen_2016} investigated the dynamic reserve prices of guaranteed contracts based on RTB. However, the model does not give the optimal solution. \cite{Hojjat_2014} discussed an idea in the allocation and serving of display ads by using predetermined fixed length streams of ads. Their framework introduces user-level perspective into the common aggregate modelling of the ad allocation problem. \cite{Zhang_2017} proposed a consumption minimization model, in which the primary objective is to minimise the user traffic consumption to satisfy all contracts. Cancellations have also been discussed in some studies, namely, the publisher can cancel a guaranteed contract later if he agrees to pay a penalty ~\citep{Babaioff_2009,Constantin_2009}.

Guaranteed contract pricing has also been discussed in several recent studies. \cite{Bharadwaj_2010} presented two algorithms to compute the price of a guaranteed contract based on the statistics of users' visits to the web pages. \cite{Najafi-Asadolahi_2014} and \cite{Fridgeirsdottir_2018_1} used queueing systems and discussed two different pricing schemes for a publisher who promises to deliver a certain number of clicks or impressions on the ads posted, where uncertain demand, traffic and click behaviour are considered.  \cite{Wang_2012_1}, \cite{Chen_2015_2} and \cite{Chen_2019_1} discussed several pricing methods for various flexible guaranteed contracts tailored to display advertising, called ad options. The ideas came from financial and real options~\citep{Constantinides_2001}. Simply, if an advertiser pays a small fee to buy an ad option, he is guaranteed a priority buying right but not an obligation of his targeted future impressions. He can then decide to pay the fixed price in the future to advertise.

Our research in this paper concerns both pricing and allocation so the optimal solution includes and reflects their interaction effects. Our problem setup is similar to~\cite{Chen_2014_2}. However, both the model and the analysis are significantly different from theirs in three important aspects. First, we consider stochastic demand for buying guaranteed contracts. We use a Poisson process to model the arrival of advertisers and allow unfulfilled demand to be backlogged. While~\cite{Chen_2014_2} assumes the demand for advertising in the future period can be shifted in advance by using a deterministic exponential decay function, the unfulfilled demand in their setting is not explicitly considered at later time points. Second, we devise an optimal pricing and allocation solution to maximise the publisher\rq{}s expected total revenue, by extending an algorithm for the Knapsack problem. Different from~\cite{Chen_2014_2} where the optimal solution is linearly searched, our solution is a greedy algorithm, and is relatively scalable and efficient. Third, we further analyse the model's robustness by incorporating supply and demand uncertainty and customising optimal pricing and allocation for different advertiser segments.

\section{Model}
\label{sec:model}

\begin{figure*}[t]
\centering
\includegraphics[width=0.7\linewidth]{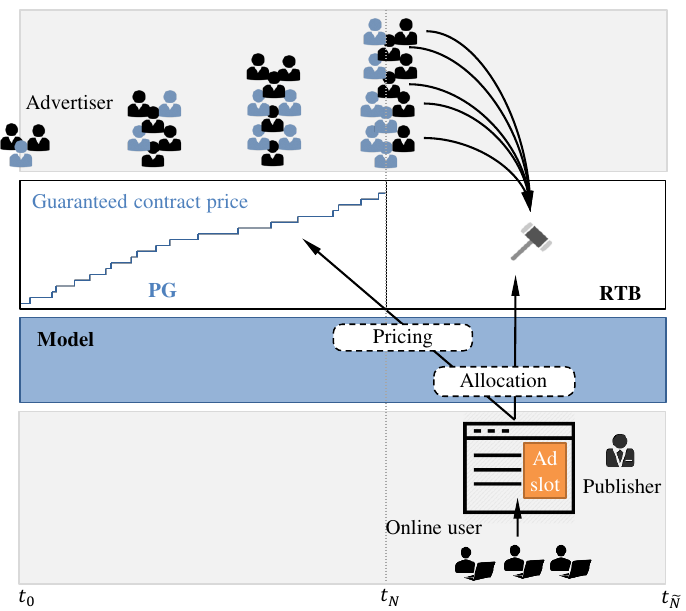}
\vspace*{7pt}
\caption{Schematic view of the proposed model for display advertising: $[t_0, t_N]$ is the time period to sell the guaranteed contracts containing the impressions which will be created in the future period $[t_N, t_{\widetilde{N}}]$; advertisers' demand of advertising in $[t_N, t_{\widetilde{N}}]$ arrives sequentially over time in $[t_0, t_N]$; and the unfulfilled demand will join RTB in $[t_N, t_{\widetilde{N}}]$.}
\label{fig:shift_of_selling}
\end{figure*}

Fig.~\ref{fig:shift_of_selling} presents a schematic view of the model. It demonstrates how a publisher can sell impressions from a specific ad slot of a publisher between guaranteed and spot markets. Specifically, impressions can be sold in advance via standardised guaranteed contracts and the remaining impressions will be auctioned off in RTB when online users visit the corresponding hosting web page. Let $[0, T]$ be the selling period of guaranteed contracts and $[T, \widetilde{T}]$ be the period that impressions are created and auctioned off in RTB. We further use $t_n$, $n = 0,1,\cdots,N$, to denote $N$ equally spaced discrete time points during the selling period. The relationship between the discrete-time and continuous-time notations is: $t_0 = 0$, $t_N = T$, and $t_{\widetilde{N}} = \widetilde{T}$. Suppose that the total supply of impressions $S$ in the future period $[t_N, t_{\widetilde{N}}]$ is well estimated, the publisher needs to decide how many future impressions to sell in advance through guaranteed contracts during $[t_0, t_N]$ and what price should be charged at each specific time point $t_n$ prior to RTB. The publisher's decision making takes the buying behaviour of advertisers into account. We assume that the total demand of future impressions $Q$ can be well estimated but advertisers arrive stochastically over time prior to the delivery day. For simplicity and without loss of generality, we assume that each advertiser has unit demand (i.e., one impression) so each guaranteed contract is a standardised contract only containing a single impression. This setting is reasonable because impressions are auctioned off individually in RTB~\citep{Ben-Zwi_2015} and the setting can be easily extended to bulk sale in practice. We assume the total demand exceeds the total supply to ensure there are competitions among advertisers in the future auctions. Otherwise, no guaranteed contracts would be purchased in advance because they can obtain the needed impressions at very low prices or even reserve prices in auctions~\citep{Yuan_2014}. For the reader's convenience, a notation table is provided in Appendix A. In our model setting, the publisher's goal is to sell $S$ impressions from a specific delivery period to $Q$ advertisers with unit demand and stochastic arrival to maximise its revenue. By allowing the publisher to sell these impressions via both guaranteed contract before the delivery period and RTB in the delivery period, we investigate two decisions that the publisher needs to make: (i) how to allocate impressions between the two channels? and (ii) how to set the unit price for the impressions sold via standardised guaranteed contracts at different time points before RTB?

\subsection{Stochastic demand arrivals and purchase behaviour}
\label{sec:stoch_demand_arrival}

A distinguishing feature of our model is that we consider stochastic demand in buying guaranteed contracts. For ease of exposition, we assume advertisers arrive following a homogeneous Poisson process with a constant~$\lambda$.\footnote{Poisson process is widely adopted in the literature mainly because of the memoryless property of the exponential inter-arrival distribution~\citep{McGill_1999,Aviv_2008}. It captures the randomness between arrivals and helps us formulate the dynamic pricing setting for the guaranteed contracts with RTB. Our model framework can also accommodate other demand arrival processes, such as non-homogeneous Poisson process. This may make the model expression more complicated but the main insights will still hold, and we believe this is a relatively minor technical concern.} Let $\Delta t = t_n - t_{n-1}$ and $f(t_n)$ be the expected arrivals in the period $\Delta t$ so $f(t_n) = \lambda \Delta t$. Once advertisers arrive, they will last up to time $t_N$ if their demand is not fulfilled, and we normalise the waiting cost to zero.\footnote{In our model, we normalise the waiting cost to zero, because the impressions are all delivered in the same time, i.e., in period $[t_N, t_{\widetilde{N}}]$. We do not consider the case where the advertisers may strategically delay their purchases to wait for a lower price in future time periods (prior to the delivery), which will require stronger assumptions regarding how the advertisers hold beliefs on future prices that they would pay for guaranteed contracts and RTB, and which cannot be observed in our dataset. We thus discuss the advertisers' strategic waiting behaviour as a future research topic in Section~\ref{sec:conclusion}.} The cumulative expected total demand of buying in advance up to time $t_N$ should not be larger than the estimated total demand in the delivery period, that is, $\lambda \leq Q/\big(\sum \Delta t \big) = Q/T$.

As the auction outcome is uncertain, we assume advertisers are risk-aware~\citep{Radovanovic_2012,Chen_2014_2} and we follow the framework proposed by~\cite{Anjos_2004,Anjos_2005} to assume that a certain percentage of arrived advertisers would like to buy guaranteed contracts. Time and price are two key factors of an advertiser's buying decision on the guaranteed contract. Therefore, a ratio function $\theta(t, p(t))$ is used to represent the proportion of those who want to buy an impression in advance at time $t$ and at price $p(t)$, satisfying the following properties:
\begin{align}
\theta(t, p)  \geq & \ \theta(t, p^*),   \hspace*{10pt} \textrm{for } 0 \leq p \leq p^*, 0 \leq t \leq T, \label{eq:demand_function_property1}\\
\theta(t, p) \geq & \ \theta(\tau, p),   \hspace*{14pt} \textrm{for } 0 \leq \tau \leq t \leq T, 0 \leq p, \label{eq:demand_function_property2}\\
\theta(t, 0) = & \ 1, 					 \hspace*{39pt} \textrm{for } 0 \leq t \leq T.\label{eq:demand_function_property3}
\end{align}
Eq.(\ref{eq:demand_function_property1}) shows that at the same time point, more advertisers are willing to buy a guaranteed contract when the price is lower; Eq.(\ref{eq:demand_function_property2}) indicates that more advertisers are willing to buy a guaranteed contract when it is closer to the end of time horizon; Eq.(\ref{eq:demand_function_property3}) denotes that all advertisers are willing to purchase a guaranteed contract when its price is zero. It is worth mentioning that the above purchase behaviour assumptions are mainly for the general rational advertisers without budget constraints. The guaranteed contract price should not exceed the advertiser's value on an impression, which will be discussed in Section~\ref{sec:time_dependent_risk_preference}.

Consistent with the existing literature, we formulate $\theta(t_n, p(t_n))$ in the following form:
\begin{align}
\theta(t_n, p(t_n)) = & \ \exp\bigg\{ - \alpha p(t_n) \Big(1 + \beta (t_N - t_n) \Big)\bigg\}, \label{eq:demand_function_1}
\end{align}
where $\alpha$ represents the price effect and $\beta$ represents the time effect. This functional form is widely used in dynamic programming with various applications, such as selling flight tickets~\citep{Anjos_2005}, display advertising~\citep{Chen_2014_2}, and so on.

Based on Eq.(\ref{eq:demand_function_1}), the demand for buying a guaranteed contract at time $t_n$ can be computed as follows:
\begin{align}
\eta(t_n) = & \
\mathbb{I}_{\{n > 0\}}
\sum_{i = 0}^{n-1} f(t_i) \prod_{j=i}^{n-1} \bigg[ 1 - \theta(t_j, p(t_j)) \bigg] + f(t_n),
\end{align}
where $\mathbb{I}_{\{\cdot\}}$ is an indicator function, $\sum_{i = 0}^{n-1} f(t_i) \prod_{j=i}^{n-1} \big (1 - \theta(t_j, p(t_j)) \big)$ computes the unfulfilled demand backlogged from the previous time periods and $f(t_n)$ is the expected number of advertisers arriving in the current time period.

\subsection{RTB-based terminal value}
\label{sec:rtb}

Another distinguished feature of our model is that, the terminal value of this dynamic programming problem depends on the outcome of RTB. In RTB, impressions are usually sold separately through the seal-bid second-price auction~\citep{Ben-Zwi_2015}, and the existing literature has shown that such a mechanism enables truth-telling, namely, it is a weakly dominant strategy to bid at one's valuation~\citep{Narahari_2014}.

Let $\xi$ be the number of advertisers who enter an RTB campaign, which can be interpreted as the competition level in RTB. By following the auction literature~\citep{Narahari_2014}, the expected revenue from RTB can be obtained as follows:
\begin{align}
\phi(\xi) = & \int_{\Omega}  x \xi (\xi - 1) g(x) \big[1 - F(x) \big] \big[F(x)\big]^{\xi-2} dx,
\label{eq:payment_exp}
\end{align}
where $x$ is an advertiser's bid, $\Omega$ is the range of bid, $g(\cdot)$ and $F(\cdot)$ are the density and cumulative distribution functions, respectively. Therefore, $\xi (\xi - 1) g(x) \big[1 - F(x) \big] \big[F(x)\big]^{\xi-2}$ represents the probability that if an advertiser who bids at $x$ is the second highest bidder, then one of $\xi - 1$ other advertisers must bid at least as much as he does and all of $\xi - 2$ other advertisers have to bid no more than he does. Usually, uniform or log-normal distributions are used for $g(\cdot)$ and $F(\cdot)$ to model the bid distribution~\citep{Ostrovsky_2011, Narahari_2014} and $\phi(\cdot)$ can be solved in closed-form if bids are uniformly distributed. However, neither distribution coincides with empirical data on many instances~\citep{Chen_2016,Chen_2014_2,Yuan_2014}. Thus, in this paper, $\phi(\cdot)$ will be learned from data and we will discuss this process in Section~\ref{sec:data}.

\subsection{Censored upper bound for pricing}
\label{sec:time_dependent_risk_preference}

When making purchase decisions, an advertiser maximises his utility by comparing the expected costs from guaranteed contract and RTB. Due to the higher risk of RTB compared to PG, the guaranteed contract price should include a risk premium which measures the uncertainty or risk that the advertiser fails to win the RTB campaign. At time $t_n$, the censored upper bound of the guaranteed contract price can be characterised as follows:
\begin{align}
\Phi(t_n) = & \ \min \bigg\{\underbrace{\phi(\xi(t_n)) + \delta(t_n) \psi(\xi(t_n))}_{:=\chi(t_n, \xi(t_n))}, \pi \bigg\},
\label{eq:censored_uppper_bound}
\end{align}
where $\pi$ is the expected maximum value of an impression and $\chi(t_n, \xi(t_n))$ is the risk-aware upper bound -- it is the sum of the expected payment in RTB and the risk premium. The risk premium is operationalised as the multiplication of the standard deviation $\psi(\xi(t_n))$ of payment prices in RTB and the advertiser's risk preference $\delta(t_n)$. Similar to $\phi(\cdot)$, $\psi(\cdot)$ and $\pi$ can be learned from data. And for $\delta(t)$, we model it with an exponential decay function $\delta(t_n) = \zeta e^{- v t_n}$ so its derivative $\delta(t)^\prime \leq \ 0$. Here, $\zeta$ represents the degree of risk aversion and $v$ represents the time effect. Similar functional forms have been widely used in asset pricing and risk analysis literature~\citep{Wilmott_2006_1}. Note that, if $N$ is large, $\delta(\cdot)$ ensures that $\chi(t_N, \xi(t_N))$ approaches $\phi(\xi(t_N))$ when the time is closer to the delivery day.

\subsection{Revenue maximisation}
\label{eq:optimisation_framework}

We next formulate the publisher's revenue maximisation problem. Let $R$ be the publisher's expected total revenue. It consists of the expected revenue from selling impressions through guaranteed contracts, denoted by $R^{PG}$, and the expected revenue from auctioning the remaining impressions in RTB, denoted by $R^{RTB}$. Thus, the revenue maximisation problem can be written as follows:
\begin{align}
\max & \ \
R =
\left\{
\begin{array}{c}
\underbrace{\sum_{n=0}^{N} (1-\omega \varpi) p(t_n) \theta\big(t_n, p(t_n)\big) \eta(t_n)}_{:= R^{PG}}\\
+ \
\underbrace{\bigg[ S - \sum_{n=0}^{N} \theta\big(t_n, p(t_n)\big) \eta(t_n) \bigg] \phi\big(\xi(t_{N})\big)}_{:= R^{RTB}}
\end{array}
\right\},
\label{eq:objective_01} \\
\textrm{s.t.}
& \ \ \
0 \leq p (t_n) \leq  \Phi (t_n), \textrm{for } n = 0, \cdots, N, \label{eq:constraint_01}\\
& \ \ \
0 \leq \sum_{n=0}^{N} \theta\big(t_n, p(t_n)\big) \eta (t_n) \leq S, \label{eq:constraint_02}
\end{align}
where $\omega$ is the probability that the publisher fails to deliver a guaranteed impression, $\varpi$ is the size of penalty proportional to the price so that the publisher needs to pay $\varpi p(t_n)$ penalty if he fails to deliver a guaranteed impression which is sold at price $p (t_n)$, and the level of competition in RTB, $\xi (t_n)$, $n = 1, \cdots, N$, can be measured by the average number of advertisers per impression as follows
\begin{align}
\label{eq:xi_n}
\xi (t_n) = & \ \frac{Q - \sum_{i=0}^{n} \theta (t_i, p(t_i)) \eta (t_i)  }{S - \sum_{i=0}^{n} \theta (t_i, p(t_i)) \eta (t_i)}.
\end{align}

In the above revenue maximisation problem, $\omega$, $\varpi$, $Q$, $S$, $N$ are treated as model parameters so their values are assumed to be given when we use dynamic programming to solve the revenue maximisation problem.\footnote{It should be noted that the advertiser pays when he buys a guaranteed contract and he can receive the penalty payment from the publisher if guaranteed impression is not delivered on the delivery day.} Eq.~(\ref{eq:constraint_01}) specifies the boundaries of the guaranteed contract price at each time point, and Eq.~(\ref{eq:constraint_02}) ensures the total amount of impressions sold via guaranteed contracts does not exceed the estimated total supply $S$ in order to prevent over-selling. The decision variable $\boldsymbol{p} = [p(t_0), \cdots, p(t_N)]$ is a vector of guaranteed contract prices which gives the pricing strategy and each $\boldsymbol{p}$ has a corresponding allocation ratio $\gamma$ representing the percentage of impressions that should be sold via guaranteed contracts. Therefore, the optimal pricing and allocation strategy can be denoted by $(\boldsymbol{p}^*, \gamma^*)$.

\subsection{Optimal solution}
\label{sec:solution}

\begin{algorithm}[tp]
\caption{$\textbf{OPT-R}$}
\label{algo:solution}
\begin{algorithmic}[1]
\State \textbf{Input:} $\alpha, \beta, \zeta, \eta, \omega, \kappa, \lambda, S, Q, T, N$
\State $\boldsymbol{t} = [t_0, \cdots, t_N]$; 	\Comment{Initialisation}
	\For{$\; n \leftarrow 0, \cdots, N$}
		\State $u_n \leftarrow \min \{S, \sum_{i=0}^{n} f(t_i)\}$;  \Comment{Sales upper bound}
		\State $l_n \leftarrow l_{n-1} \ \mathbb{I}_{\{n > 0\}}$; \Comment{Sales lower bound}
		\State $\mathcal{Y}_n \leftarrow \{l_n, l_n+1,\cdots, u_n\}$;\Comment{Set of sales at time $t_n$}		
		\For{$\; y \in \mathcal{Y}_n$}
			\State $\mathbf{H}_n(y) \leftarrow \textbf{OPT-H}(\alpha, \beta, \zeta, \eta, \omega, \kappa, \lambda, S, Q, T, n, y)$; \Comment{Algorithm~\ref{algo:calc_hn}}
			\If{$\; n = N$}
				\State $R(y) \leftarrow \mathbf{H}_n(y) + (S - y) \phi\Big(\frac{Q-y}{S-y}\Big)$;
			\EndIf
		\EndFor
	\EndFor
\State \textbf{Output:} $R^* \leftarrow \max_{y \in \mathcal{Y}_N} \{R(y)\}$; $\{\mathbf{H}_n(y): n = 0, \ldots, N; y \in \mathcal{Y}_n\}$
\end{algorithmic}
\end{algorithm}

\begin{algorithm}[t]
\caption{$\textbf{OPT-H}$}
\label{algo:calc_hn}
\begin{algorithmic}[1]
\State \textbf{Input:} $\alpha, \beta, \zeta, \eta, \omega, \kappa, \lambda, S, Q, T, n, y, \mathbf{H}_{n-1}$
\State $\boldsymbol{z} \leftarrow$ an allocation matrix with size $(y+1) \times 2$; 	\Comment{Initialisation}
\State $j \rightarrow 1$ \Comment{Index initialisation of $\widetilde{\mathbf{H}}$}
\For{$\; k \leftarrow 1, \cdots, (y+1)$}
	\State $p_{n,j,k} \leftarrow $ Eq.~(\ref{eq:guaranteed_price});
	\State $\Phi_{n,j,k} \leftarrow $ Eq.~(\ref{eq:censored_uppper_bound});
	\If{$p_{n,j,k} > \Phi_{n,j,k}$} \Comment{Check price bound}
		\State \textrm{Continue};
	\EndIf
	\State $\widetilde{\mathbf{H}}_{n,j}(y) \leftarrow \mathbb{I}_{\{n>0\}}\mathbf{H}_{n-1}(\boldsymbol{z}(k,1)) + p_{n,j,k} \boldsymbol{z}(k,2)$;
	\State $j \rightarrow j+1$
\EndFor
\State \textbf{Output:} $\mathbf{H}_n(y) \leftarrow \max_j\{ \widetilde{\mathbf{H}}_{n,j}(y)\}$;
\end{algorithmic}
\end{algorithm}

\begin{figure}[t]
\centering
\includegraphics[width=1\linewidth]{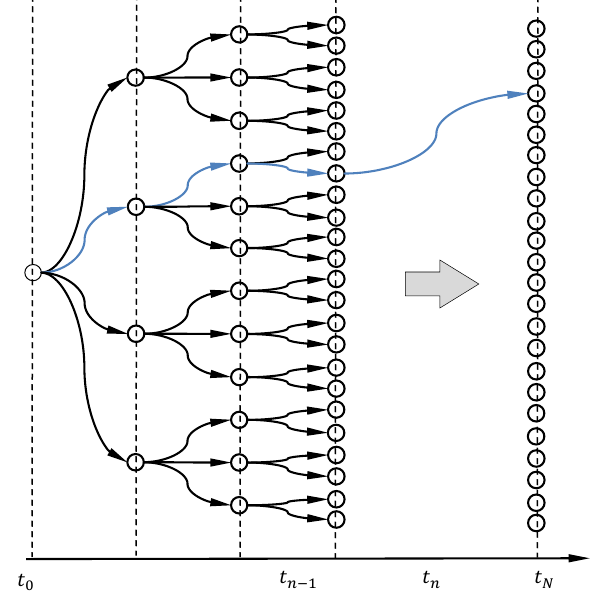}
\caption{Schematic view of the proposed optimal solution. Each node represents a possible scheme of allocation and pricing of impressions at the corresponding time steps.}
\label{fig:optimal_solution}
\end{figure}

The main challenge of solving the publisher's revenue maximisation problem is that $R^{RTB}$ is uncertain because the action of selling guaranteed contracts will affect both the supply and demand of impressions in RTB, which further impacts its expected revenue. Our problem is similar to the Knapsack problem~\citep{Kleinberg_2005}, where the optimisation deals with a sequence of items with values and non-negative weights. In our problem, prices are affected by the allocation and they are further censored based on the buying behaviour of advertisers. As the expected total revenue can be divided into $R^{PG}$ and $R^{RTB}$, when the remaining total supply and demand are given, $R^{RTB}$ can be estimated consequently. So the optimal $R^{PG}$ can be obtained using the recurrent structure of dynamic programming. Our solution is presented in Algorithms~\ref{algo:solution}-\ref{algo:calc_hn}. Algorithm~\ref{algo:solution} initiates advertisers' arrivals and computes the optimal total expected revenue based on different allocation schemes. For each allocation scheme, Algorithm~\ref{algo:calc_hn} computes the optimal expected revenue of selling the guaranteed contracts. Fig.~\ref{fig:optimal_solution} presents a schematic view of our solution. In essence, we create a decision tree over time, in which each node represents a possible scheme of allocation and the corresponding price at a specific time point. The optimal allocation and pricing strategy is the trial over the tree which maximises the expected total revenue. Next, we explain a few key steps of both algorithms in detail.

As the demand for buying guaranteed contracts arrives following a Poisson process, together with Eq.~(\ref{eq:constraint_02}), the upper bound $u_n$ of total amount of sold impressions up to time $t_n$ can be defined by $\min \{S, \sum_{i=0}^{n} f(t_i)\}$. Its lower bound $l_n$ is given from time $t_{n-1}$ and $l_0$ is zero. The possible sold impressions up to time $t_n$ can then be generated, denoted by set $\mathcal{Y}_n$. For any $y \in \mathcal{Y}_n$, its optimal subset-sum can be computed by creating a $(y+1) \times 2$ matrix $\boldsymbol{z}$, which contains all possible combinations of impressions sold up to time $t_{n-1}$ and at time $t_n$, such that $\boldsymbol{z}(k,1) + \boldsymbol{z}(k,2) = y$, for $k = 1, \cdots, y+1$, and $\boldsymbol{z}(k,1) = 0$ if $n=0$. Since
\begin{align*}
\frac{
\boldsymbol{z}(k,2)
}{
\sum_{i=0}^{n} f(t_i) - \boldsymbol{z}(k,1)
}
\propto & \ \theta(t_n, p(t_n)),
\end{align*}
the price of a guaranteed contract at time $t_n$ can be obtained as
\begin{equation}
\label{eq:guaranteed_price}
p_{n,j,k} = - \frac{\ln\{z(k,2)\} - \ln\Big\{ \sum_{i=0}^{n} f(t_i) - z(k,1) \Big\}}{\alpha \big(1+\beta (t_N - t_n)\big)}.
\end{equation}
We then obtain $\Phi_{n,j,k}$ by Eq.~(\ref{eq:censored_uppper_bound}). If $p_{n,j,k} > \Phi_{n,j,k}$, the price $p_{n,j,k}$ will be removed from the solution space and this price will be replaced with the value calculated by the next index of $k$. If $p_{n,j,k} \leq \Phi_{n,j,k}$, we can calculate the corresponding revenue of selling guaranteed contracts up to time $t_n$, denoted by $\widetilde{\mathbf{H}}_{n,j}(y)$. Then, the maximised revenue of selling guaranteed contracts up to time $t_n$ is obtained, denoted by $\mathbf{H}_n(y)$. The iterations carry on until time $t_N$. For each $y \in \mathcal{Y}_N$, the expected total revenue can be obtained by adding $\mathbf{H}_n(y)$ with the corresponding expected revenue from RTB. Finally, the optimal solution can be obtained by comparing the expected total revenues from all candidates $\{R(y): y \in \mathcal{Y}_N\}$ in the solution space. As the optimal selling amount and the corresponding price of each time step have been stored, we can use the index of the maximised revenue to obtain the optimal price vector $\boldsymbol{p}^*$ and the corresponding allocation $\gamma^*$. Our solution is a greedy algorithm and the time complexity is $\mathcal{O}(N S^2)$, where $\mathcal{O}$ is a notation used to classify algorithms according to how the running time grows as the input size grows. As $N$ is much less than $S$, our solution is relatively scalable and efficient.

\begin{figure*}
\centering
\includegraphics[width=1\linewidth]{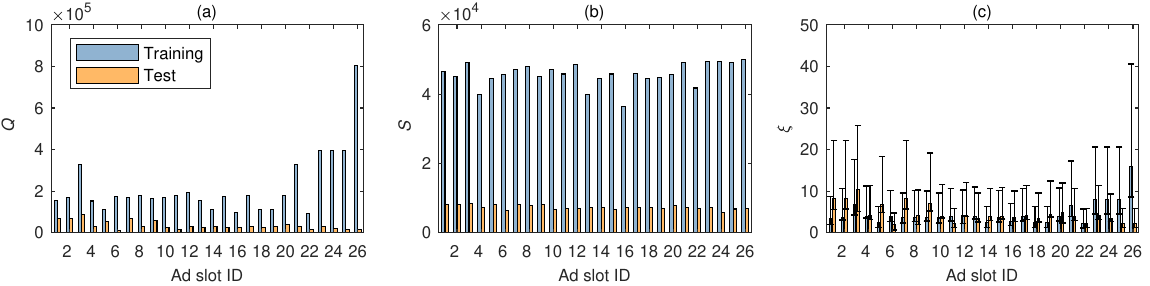}
\caption{Summary of the RTB dataset: (a) the total demand; (b) the total supply; (c) the average per impression competition level.}
\label{fig:demand_supply}
\end{figure*}

\begin{table*}[t]
\centering
\caption{Summary of clustered ad slots, where numbers in round brackets are standard deviations.}
\label{tab:stats_groups}
\begin{tabular}{r|c|c|c|c}
\toprule
Group
& \multicolumn{2}{c|}{1}
& \multicolumn{2}{c}{2}\\
\hline
Number of ad slots
& \multicolumn{2}{c|}{6}
& \multicolumn{2}{c}{20}\\
\hline
Set
& Training
& Test
& Training
& Test\\
\hline
Payment price	& 0.98 (0.09) & 0.99 (0.08) & 0.73 (0.46) & 0.56 (0.36)\\
\hline
Winning bid	& 1.13 (0.17) & 1.1 (0.1) & 2.32 (1.17) & 1.84 (1.04)\\
\hline
$\xi$ 		& 8.92 (3.24) & 8.15 (1.18) & 3.39 (0.59) & 3.51 (0.81)\\
\hline
Ratio of payment		& 88.95\% 	& 92.88\% 	& 32.2\% 	& 37.18\%	\\
price to winning bid    & (4.54\%) & (2.15\%) & (9.9\%) 	& (10.58\%)\\
\bottomrule
\end{tabular}
\end{table*}

\begin{figure*}[t]
\centering
\includegraphics[width=0.7\linewidth]{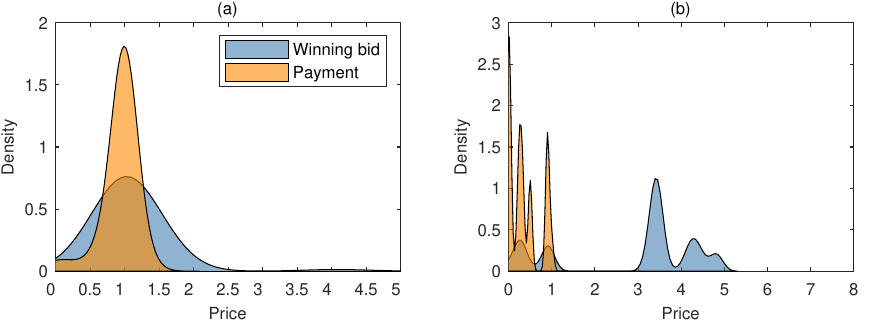}
\caption{Examples of distributions of winning bids and payment prices in the training set: (a) ad slot 24 in group 1; (b) ad slot 15 in group 2.}
\label{fig:example_price_dist}
\end{figure*}

\section{Data and experimental settings}
\label{sec:data}

We use an RTB dataset from a UK SSP to validate the proposed model of selling impressions via both guaranteed contracts and RTB.\footnote{The dataset has also been used in several recent display advertising studies~\citep{Chen_2014_2,Yuan_2014,Chen_2016}.} This dataset contains 1,378,971 RTB campaigns for 31 different ad slots over the period from 08 January 2013 to 14 February 2013. For each ad slot, RTB campaigns range from 7 to 20 continuous days. Therefore, we select the campaign records of 7 continuous days from all ad slots for experiments. Given an ad slot, the delivery day $[t_N, t_{\widetilde{N}}]$ is randomly selected and the campaigns reported from this day is used for validation and testing (called the \emph{test set}). The records of continuous 6 days prior to the delivery day are used to estimate the model parameters (called the \emph{training set}). For those ad slots with only 7 days data, the 7th day is set to be the delivery day. In our dataset, all bids in RTB campaigns are quoted in terms of cost-per-mille (CPM), which is a measurement corresponds to the value of 1,000 impressions~\citep{YYuan_2014}.

The total supply $S$ and demand $Q$ of impressions for a specific ad slot in the delivery day can be predicted by time series or regression models. As our primary intention here is not to discuss a prediction model, $Q$ and $S$ are simply given by the test set. Fig.~\ref{fig:demand_supply} summarises the total supply and demand of impressions for ad slots in both training and test sets. There are 5 slots excluded from the original dataset because the competition level of the slot in RTB is less than 2 in the training set (i.e., $\xi < 2$). In such cases, advertisers are able to obtain the needed impressions in RTB at a very low price or even the reserve price. Therefore, it is very unlikely for these advertisers to buy any guaranteed contracts in advance. Overall, it appears that the total supply levels are similar across ad slots while the total demand levels are significantly different. It is worth noting that, within a day, the competition level $\xi$ varies significantly from its mean as many advertisers join RTB at peak hours between 6 am to 10 am, making $\xi$ on average 118.96\% higher than other hours. In the experiments, we assume guaranteed contracts can be sold 31 days prior to the delivery day, and as discussed in Section~\ref{sec:model}, the advertiser arrivals follow a Poisson process during the 31 days. To simplify the discussion and without loss of generality, intensity $\lambda$ is set as a constant. Specifically, 20\% of $Q$ is assumed to arrive at time 0 and another 20\% of $Q$ is considered to arrive prior to the delivery day. Therefore, $\lambda = 0.2 Q/30$.\footnote{Our results with different values for $\lambda$ show that, consistent with intuition, smaller (larger) $\lambda$ leads to less (more) guaranteed contracts being sold out. However, the main insights from the analysis remain the same.}

We divide 26 ad slots into two groups using the K-Means clustering~\citep{Bishop_2006} based on the competition level $\xi$ in the training set. Tables~\ref{tab:stats_groups} presents a brief summary of the major statistics of both groups. The table shows that, group 1 with a high competition level has the average 8.92 (in the training set) and 8.15 (in test set) advertisers bidding per RTB auction, much higher than those of group 2 with a low competition level. Consistent with Eq.~(\ref{eq:payment_exp}), the average of per-auction payments of group 1 are also higher than in group 2. Interestingly, group 2 has a higher average winning bid. By further investigating the distributions of winning bids and payment prices, we find that the counter-intuitive result is due to the different price patterns over two groups, which is demonstrated in Fig.~\ref{fig:example_price_dist}. Fig.~\ref{fig:example_price_dist} uses two examples, slot 24 in group 1 and slot 15 in group 2. For slot 24, the distributions of winning bids and payment prices are bell shaped, and payment price has a higher peak and a slightly smaller mean. Differently, for ad slot 15, the distributions of winning bids and payment prices are similar to Gaussian mixture distributions~\citep{Bishop_2006}, and payment prices are much lower than winning bids. The ratio of payment price to winning bid for group 1 is 88.95\% while it is 32.2\% for group 2. This suggests that, for ad slots in the group with a low competition level, there is greater potential to optimise the selling mechanism to increase the revenue. Further comparing the training set and the test set in both group 1 and 2, we observe from Table~\ref{tab:stats_groups} that most statistics are close and consistent. Thus, it makes sense that we use the training set to estimate the model parameters, and validate the model in the test set.

\begin{figure*}[t]
\centering
\includegraphics[width=1\linewidth]{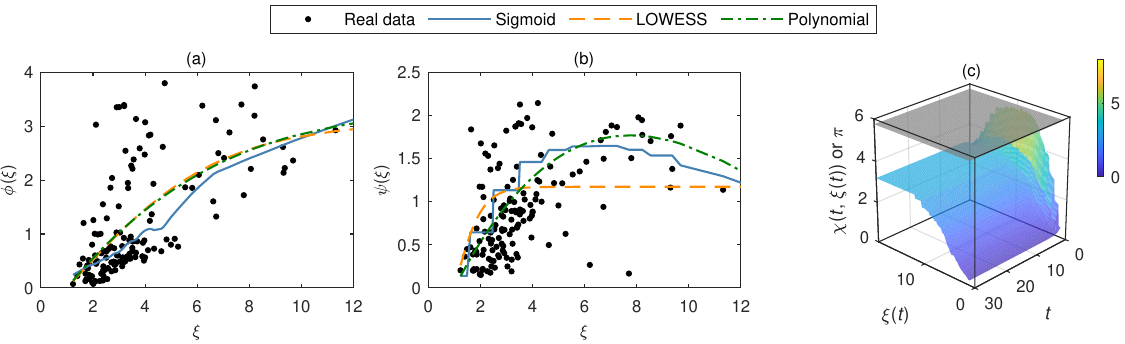}
\caption{Examples of estimating the censored price upper bound of guaranteed contracts for ad slot 6 in group 2 based the training data (hourly basis): (a) $\phi(\xi)$; (b) $\varphi(\xi)$; and (c) surfaces of $\chi(t, \xi(t))$ and $\pi$ (in gray colour).}
\label{fig:example_price_fit}
\end{figure*}

We further use the dataset to estimate parameters $\phi(\xi)$, $\varphi(\xi)$ and $\chi(t, \xi(t))$ when $\xi$ is given, which are visualised in Fig.~\ref{fig:example_price_fit}. Following existing literature~\citep{Chen_2014_2,Chen_2016}, we implement the locally weighted regression scatterplot smoothing (LOWESS), polynomial regression, and sigmoid methods.\footnote{Other statistical or machine learning methods may also be used here, however, discussing the best prediction method is out of the scope of this paper.} The LOWESS method combines multiple weighted polynomial regression models~\citep{Cleveland_1979} and we follow the implementation settings of Algorithm~1 given in~\cite{Chen_2014_2}. The sigmoid method uses a scaled sigmoid function to approximate the training data, where the scaling parameters are calibrated based on the root mean squared errors. Finally, the LOWESS method is selected for parameters estimation as it fits the training data best. Given time $t$, $\delta(t)$ is calculated and then $\chi(t, \xi(t))$ is obtained. We set $\pi$ as the maximum value of average bids (hourly basis) in the training set, then a surface of the price upper bound $\Phi(t)$ is obtained.

\section{Results}
\label{sec:empirical_results}

In this section, we first present our experimental results on the model's performance, and then investigate the model's robustness by considering: (i) the effect of uncertainty in supply and demand of impressions; and (ii) advertisers with different valuations. The robustness analysis can provide insights for daily operations as the two situations may occur in practice.\footnote{We thank one of the reviewers for suggesting the extensions in Sections~\ref{sec:incorporating_uncertainty}-\ref{sec:advertiser_segmentation}.}

\subsection{Optimal pricing and allocation}
\label{sec:optimal_pricing_and_allocation}

\begin{figure*}[t]
\centering
\includegraphics[width=1\linewidth]{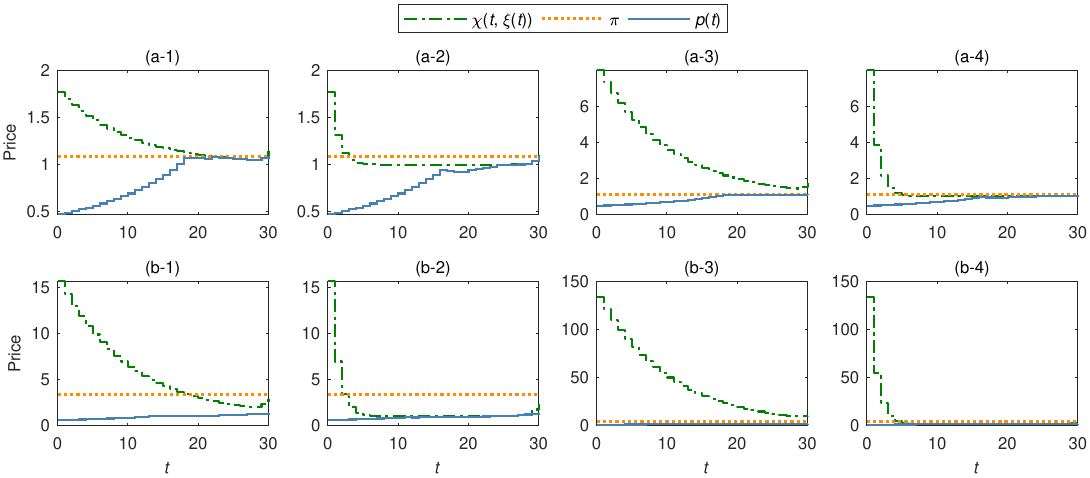}
\caption{Examples of optimal pricing of guaranteed contracts suggested by the model for: (a) ad slot 24 in group 1 and (b) ad slot 15 in group 2. Parameters are set differently in the subplots: (1) $\zeta = 10, v = 0.1$; (2) $\zeta = 10, v=0.9$; (3) $\zeta = 90, v = 0.1$; (4) $\zeta = 90, v = 0.9$.}
\label{fig:example_price_movement}
\end{figure*}

\begin{figure*}[t]
\centering
\includegraphics[width=1\linewidth]{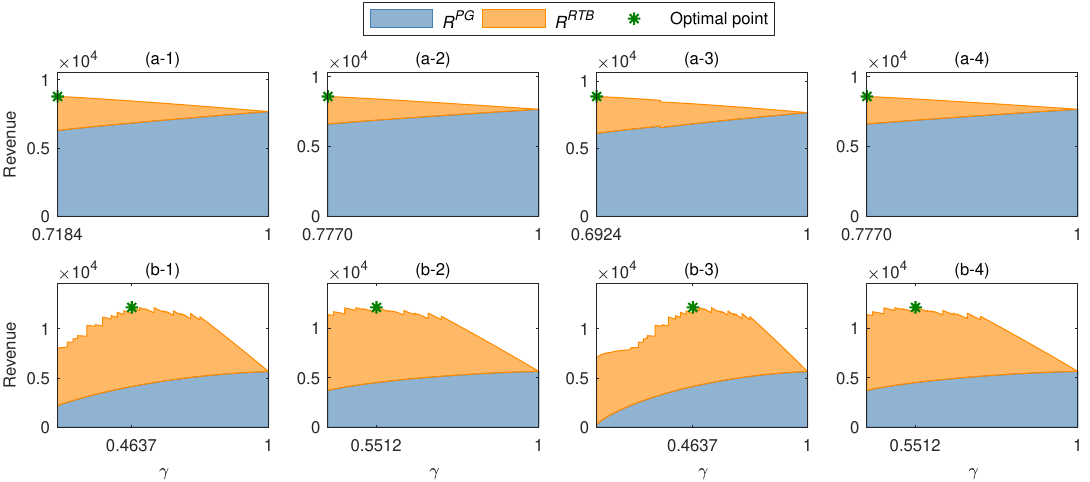}
\caption{Examples of optimal allocation of guaranteed contracts suggested by the model for: (a) ad slot 24 in group 1 and (b) ad slot 15 in group 2. Parameters are set differently in the subplots: (1) $\zeta = 10, v = 0.1$; (2) $\zeta = 10, v=0.9$; (3) $\zeta = 90, v = 0.1$; (4) $\zeta = 90, v = 0.9$.}
\label{fig:example_allocation}
\end{figure*}

\begin{figure*}[t]
\centering
\includegraphics[width=1\linewidth]{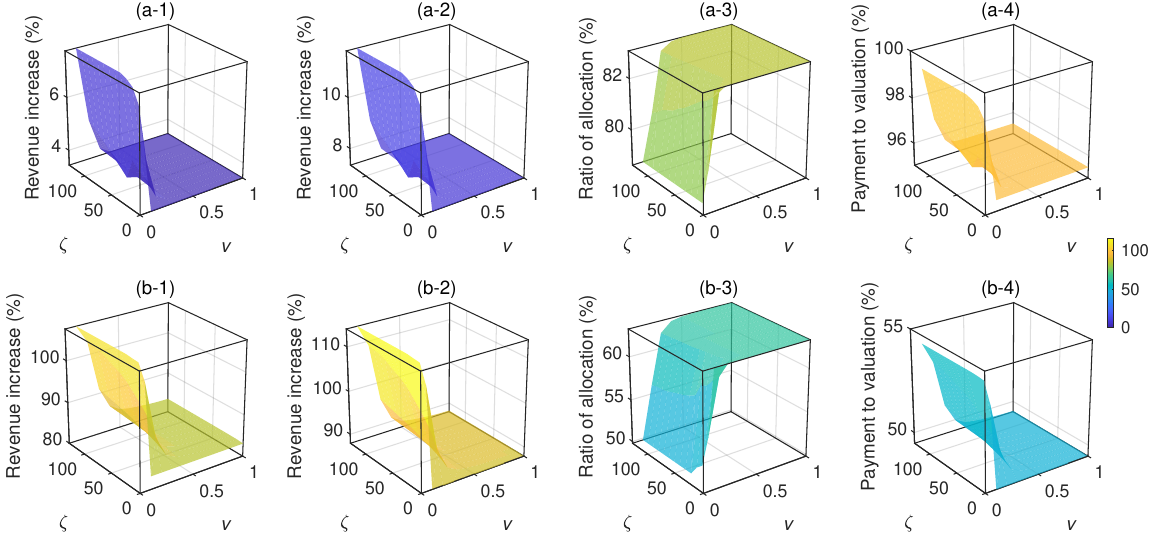}
\caption{Overall results of the model performance on ad slots in: (a) group 1; (b) group 2. The sub-plots show: (1) the average revenue increase of the model to the expected RTB; (2) the average revenue increase of the model to the actual RTB; (3) the average ratio of selling impressions in advance made by the model; (4) the average ratio of payment to valuation made by the model.}
\label{fig:overall_results}
\end{figure*}

To illustrate how the proposed model can optimally price and allocate impressions between guaranteed and spot markets, Figs.~\ref{fig:example_price_movement}-\ref{fig:example_allocation} present examples of ad slot 24 from group 1 and ad slot 15 from group 2. We deliberately show the results of these two slots because they are typical instances in the corresponding group. Also, as the distributions of wining bids and payment prices of both slots in RTB have been shown in Fig.~\ref{fig:example_price_dist}, we can clearly see if the optimal prices of guaranteed contracts are feasible.

Fig.~\ref{fig:example_price_movement} illustrates the optimal pricing strategies suggested by the model as well as the effects of the advertisers' risk preference (i.e., $\zeta$ and $v$) on the optimal prices of guaranteed contract. Comparing the sub-figures in the same column, we can see that there is a general trend of price increase over time for both slots. The price on slot 15 increases more steadily. This is because there is less demand for buying guaranteed contracts prior to the delivery day due to the low competition level. Also, as less advertisers arrive, the price growth is not significant over time. On the contrary, as it is difficult to obtain impressions in RTB for slot 24 with high competition level, more advertisers would be willing to secure the impressions in advance through guaranteed contracts. Consequently, the guaranteed contracts can be sold at higher prices. This is confirmed by the observation in Fig.~\ref{fig:example_price_movement} that the price increases steeply and then be censored by the upper bound $\Phi(t)$, which is the minimum value between $\pi$ and $\chi(t, \xi(t))$, as defined in Eq.~(\ref{eq:censored_uppper_bound}). Furthermore, by comparing the sub-figures in the same row, Fig.~\ref{fig:example_price_movement} shows how the advertisers' risk preference parameters $\zeta$ and $v$ affect $\chi(t, \xi(t))$ and the optimal price. The value of $\chi(t, \xi(t))$ increases significantly with the increase of $\zeta$. Since $\chi(t, \xi(t))$ is an exponential decay function of $v$, it converges quickly to $\phi(\xi(t))$ if $v$ increases. The optimal price is more sensitive to $v$. However, the impacts of $v$ and $\zeta$ on the optimal price is mild.

We next investigate the optimal allocation in Fig.~\ref{fig:example_allocation}. This figure shows the details of the corresponding optimal allocations made by the model for the same ad slot and under the same experimental settings of Fig.~\ref{fig:example_price_movement}. Obviously, the model suggests different allocation strategies for different slots. For slot 24 with high competition level, the model allocates more future impressions into guaranteed contracts, and they constitute a large percentage of the expected total revenue. For slot 15 with low competition level, the model allocates only a small amount of impressions in advance. As a result, the average payment price of impressions auctioned in RTB increases significantly, and the expected total revenue is largely contributed by RTB rather than PG. Furthermore, Fig.~\ref{fig:example_allocation} also indicates the effects of advertisers' risk preference, $\zeta$ and $v$, on optimal allocation. First, the proportion of impressions allocated to PG, $\gamma$, is decreasing in $\zeta$. This is because guaranteed contract prices are slightly higher if $\zeta$ is large so less advertisers would be willing to buy guaranteed contracts. Second, $\gamma$ is increasing in $v$. This is because if $v$ increases: (i) advertisers are more risk averse and more sensitive to time; (ii) $\chi(t, \xi(t))$ decreases quickly and the guaranteed prices will be slightly lower so more advertisers would be willing to buy in advance.\footnote{The optimal allocations are same in Fig.~\ref{fig:example_allocation} (a-2) and (a-4) as well as in Fig.~\ref{fig:example_allocation} (b-2) and (b-4). This because the effects of $\zeta$ and $v$ are not significant in the chosen ad slots under our experimental settings. In Appendix B, we provide two additional examples to further justify our analysis.}

The overall results of the model performance on ad slots of two clustered groups are summarised in Fig.~\ref{fig:overall_results}. We compare the expected total revenue given by the model with both the expected RTB revenue and the actual RTB revenue. The expected RTB revenue is calculated by multiplying the average per-auction payment with the total number of impressions, while the actual RTB revenue is the total revenue reported by actual RTB campaigns in the test set.\footnote{This is also called \emph{ground truth} in machine learning literature.} Our results show that selling impressions via both PG and RTB indeed increase the total revenue for two groups with different competition levels. With the increase of $\zeta$, the increase in revenue converges to its maximum amount. Furthermore, by comparing the allocation for group 1 and 2, more impressions are allocated into guaranteed contracts in group 1 when the competition level is high, consequently, the expected total revenue is largely contributed by guaranteed contracts. However, as mentioned previously in Table~\ref{tab:stats_groups}, there is greater potential to improve revenue in ad slots with low competition level,  as its average ratio of payment price to winning bid is only around 30\%, compared to 90\% for the ad slots with high competition level. Therefore, if the ratio of payment to valuation\footnote{The ratio of payment to valuation is equivalent to the ratio of payment to winning bid in Table~\ref{tab:stats_groups}.} increases for the ad slots with low competition level, the revenue increase will be more significant, and our numerical analysis shows that in some cases the expected total revenue can even be doubled.

\subsection{Incorporating uncertainty into model updating}
\label{sec:incorporating_uncertainty}

\begin{figure*}[t]
\centering
\includegraphics[width=0.7\linewidth]{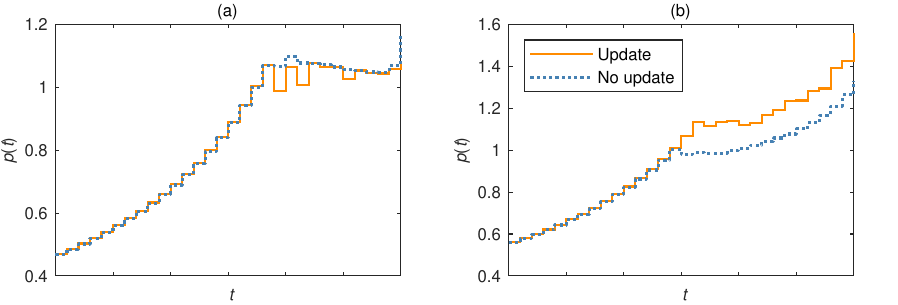}
\caption{Examples of optimal pricing of guaranteed contracts suggested by the model by adding 10\% uncertainty level ($\epsilon=0.1$) for: (a) ad slot 24 in group 1 and (b) ad slot 15 in group 2. Parameters $\zeta = 10, v = 0.1$.}
\label{fig:example_price_movement_uncertainty}
\end{figure*}

We next further investigate the model by considering the effect of uncertainty in supply and demand. In the model setting, given a specific ad slot, $S$ and $Q$ are assumed to be well estimated by the forecasting models at the current time $t_0$. However, they may change over time due to uncertainty.\footnote{If we assume they move either up and down separately, their behaviour over time can be captured by two multi-step binomial tree or lattice models~\citep{Cox_1979,Chen_2014_2} and there will be four different combinations of their estimations at each time step. If they can either move upwards, downwards, or stay unchanged in a given time period, their behaviour can be modelled by two multi-step trinomial tree or lattice models~\citep{Boyle_1986} and there will be eight different combinations of their estimations at each time step.} As is shown in Eq.(\ref{eq:xi_n}), the uncertainty from both the supply or demand will finally impact the model performance through $\xi$ the per-impression competition level in RTB. For model tractability and ease of exposition, we can fix $S$ and just consider the random behaviour of $Q$ over time.\footnote{The supply of impressions are usually stable with relatively low uncertainty, as the traffic or the number of visitors to the web pages are usually predictable. For example, tools like Google Analytics (\url{https://analytics.google.com}) and SimilarWeb (\url{https://www.similarweb.com}) offer detailed website traffic estimation service. \cite{Lee_1999,Ilfeld_2002} have also shown that many popular websites have rather stable traffic over a short-term period such as week or month.} Let $Q_{n+1}$ be the rest of total demand in $[t_N, t_{\widetilde{N}}]$ estimated at time $t_{n+1}$ and let $Q_{n}$ be the rest of total demand estimated at time $t_{n}$. We can simply use a random walk structure to update uncertainty in $Q$ as follows: $Q_{n+1} \leftarrow Q_{n} (1 + \epsilon \varepsilon)$, where $\epsilon$ is the uncertainty level expressed as a percentage and $\varepsilon$ is a white noise such that $\mathbb{E}[\varepsilon] =0$ and $\mathrm{var}[\varepsilon]=1$. In experiments, we set $\epsilon = 10\%$. When the time step moves, the sold impressions are then removed from $S$ and $Q$ accordingly. We then generate a new estimate for the rest of demand, update the corresponding model parameters, and recompute the optimal prices and allocations of guaranteed contracts for the rest of the days in the selling period. The iterations will continue until time $t_N$. By updating the uncertainty, the time complexity becomes $\mathcal{O}(N^3 S^2)$.

Fig.~\ref{fig:example_price_movement_uncertainty} compares the optimal prices with and without considering the uncertainty for ad slot 24 in group 1 and ad slot 15 in group 2. The results show that uncertainty affects the optimal prices marginally. The increasing trend of the price over time remains the same, although there are small fluctuations in price movement. Table~\ref{tab:uncertainty_groups} summarises how uncertainty impacts the optimal allocation and the expected revenue. The results show that that uncertainty indeed impacts the optimal impression allocation between PG and RTB and their respective expected revenues. Such impacts are marginal on the expected revenue, but can change the allocation significantly.

\begin{table*}[t]
\centering
\caption{Summary of changes of model allocation and revenue for clustered ad slots, where numbers in round brackets are standard deviations, where $\epsilon = 10\%$.}
\label{tab:uncertainty_groups}
\begin{tabular}{c|c|c|c|c|c|c}
\toprule
\multirow{2}{*}{Group} & \multicolumn{2}{c|}{Setting} & \multicolumn{4}{c}{Changes (\%)}\\
\cline{2-7}
	  &   $\zeta$ & $v$   & $\gamma$    & $R^{RTB}$ $\dagger$ & $R^{PG}$	& $R$ \\
\toprule	
\multirow{4}{*}{1}	
	  &	10 & 0.1 &8.29 (4.44)&-19.59 (6.16)&6.64 (4.28)&-1.15 (0.46)\\
 	  &	10 & 0.9 &2.54 (2.97)&-6.84 (10.27)&3.16 (3.43)&0.38 (0.3)\\
 	  &	90 & 0.1 &37.72 (19.21)&-41.77 (6.79)&23.88 (12.2)&-3.58 (0.55)\\
 	  &	90 & 0.9 &2.54 (2.97)&-6.84 (10.27)&3.16 (3.43)&0.38 (0.3)\\
\hline	
\multirow{4}{*}{2}	
	  &	10 & 0.1 &-8.74 (13.94)&-4.33 (23.76)&-4.3 (9.4)&-3.97 (14.44)\\
 	  &	10 & 0.9 &-6.71 (15.18)&-1.75 (17.97)&-4.59 (9.38)&-1.81 (8.96)\\
 	  &	90 & 0.1 &-12.72 (16.71)&-4.03 (24.15)&-6.05 (10.4)&-4.94 (13.95)\\
 	  &	90 & 0.9 &-6.71 (15.19)&-1.74 (17.99)&-4.55 (9.39)&-1.79 (8.96)\\
\bottomrule
\multicolumn{7}{p{5.5in}}{$\dagger$ Ad slot 26 in group 1 is excluded in the computation as the model suggests all impressions to be sold in advance via guaranteed contracts so its RTB revenue is 0.}\\
\end{tabular}
\vspace*{15pt}
\caption{Summary of clustered subgroups of ad slots, where numbers in round brackets are standard deviations. }
\label{tab:stats_groups_clustering}
\begin{tabular}{r|c|c|c|c|c}
\toprule
\multicolumn{2}{c|}{} &
\multicolumn{4}{c}{Set}\\
\cline{3-6}
\multicolumn{2}{c|}{}
& \multicolumn{2}{c|}{Training}
& \multicolumn{2}{c}{Test}\\
\cline{2-6}
& \diaghead{\theadfont Group Subgroup}{Group}{Subgroup}&
\thead{1}&\thead{2}&\thead{1}&\thead{2}\\
\toprule
\multirow{4}{*}{Payment price}
& \multirow{2}{*}{1}  & 1.17   & 0.65   & 1.16  	& 0.62 \\
&	& (0.12) & (0.34) &	(0.12) 	& (0.4)\\
\cline{2-6}	
& \multirow{2}{*}{2} & 1.09   & 0.16   & 1.03 	& 0.2 \\
&	& (0.61) & (0.07) &	(0.62)  & (0.09)\\
\hline
\multirow{4}{*}{Winning bid}
& \multirow{2}{*}{1} & 2.63 	 & 1.0 		& 2.62 	 & 0.99  \\
& 	& (1.45) & (0.04)   & (1.49) & (0.03)\\
\cline{2-6}	
& \multirow{2}{*}{2} & 3.25   & 0.62   & 3.36   & 0.53 \\
&   & (1.48) & (0.22) & (2.03) & (0.2)\\
\hline
\multirow{4}{*}{$\xi$}
& \multirow{2}{*}{1} & 14.39  & 5.94  & 14.86 	& 5.53 \\
&   & (5.26) & (2.7) & (6.39)   & (2.91)\\
\cline{2-6}	
& \multirow{2}{*}{2} & 4.1   & 2.61   & 5.17   & 2.71 \\
& 	&(1.04) & (0.48) & (2.25) & (0.74)\\
\hline
& \multirow{2}{*}{1} & 61.98\%   & 63.16\%   & 63.69\%   & 61.59\%  \\
Ratio of payment
&   & (31.56\%) & (32.63\%) & (34.23\%) & (39.75\%)\\
\cline{2-6}	
price to winning bid
& \multirow{2}{*}{2} & 38.15\%   & 28.4\%    & 42.51\%   & 35.07\%  \\
&   & (19.19\%) & (10.33\%) & (25.96\%) & (11.69\%)\\
\bottomrule
\end{tabular}
\end{table*}

\begin{figure*}[t]
\centering
\includegraphics[width=0.85\linewidth]{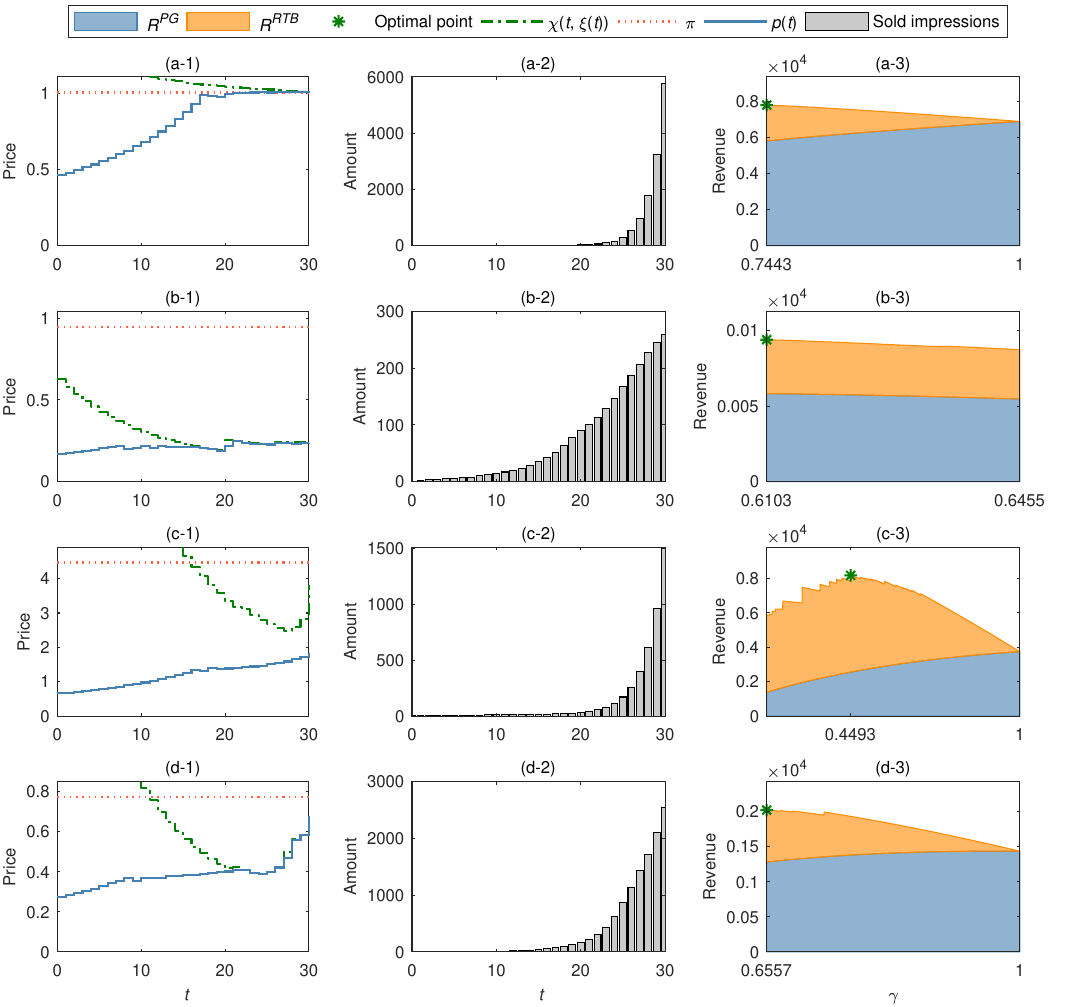}
\caption{Examples of optimal pricing, allocation and selling of guaranteed contracts suggested by the model. Sub-plots: (a) subgroup 1 of ad slot 3 in group 1; (b) subgroup 2 of ad slot 3 in group 1; (c) subgroup 1 of ad slot 15 in group 2; and (d) subgroup 2 of ad slot 15 in group 2. Sub-plots: (1) guaranteed prices and the boundaries over time; (2) sold guaranteed impressions over time; (3) expected total revenue and the optimal allocation. Parameters are set $\zeta = 10, v = 0.1$.}
\label{fig:example_clustering}
\end{figure*}

\begin{figure*}[t]
\centering
\includegraphics[width=1\linewidth]{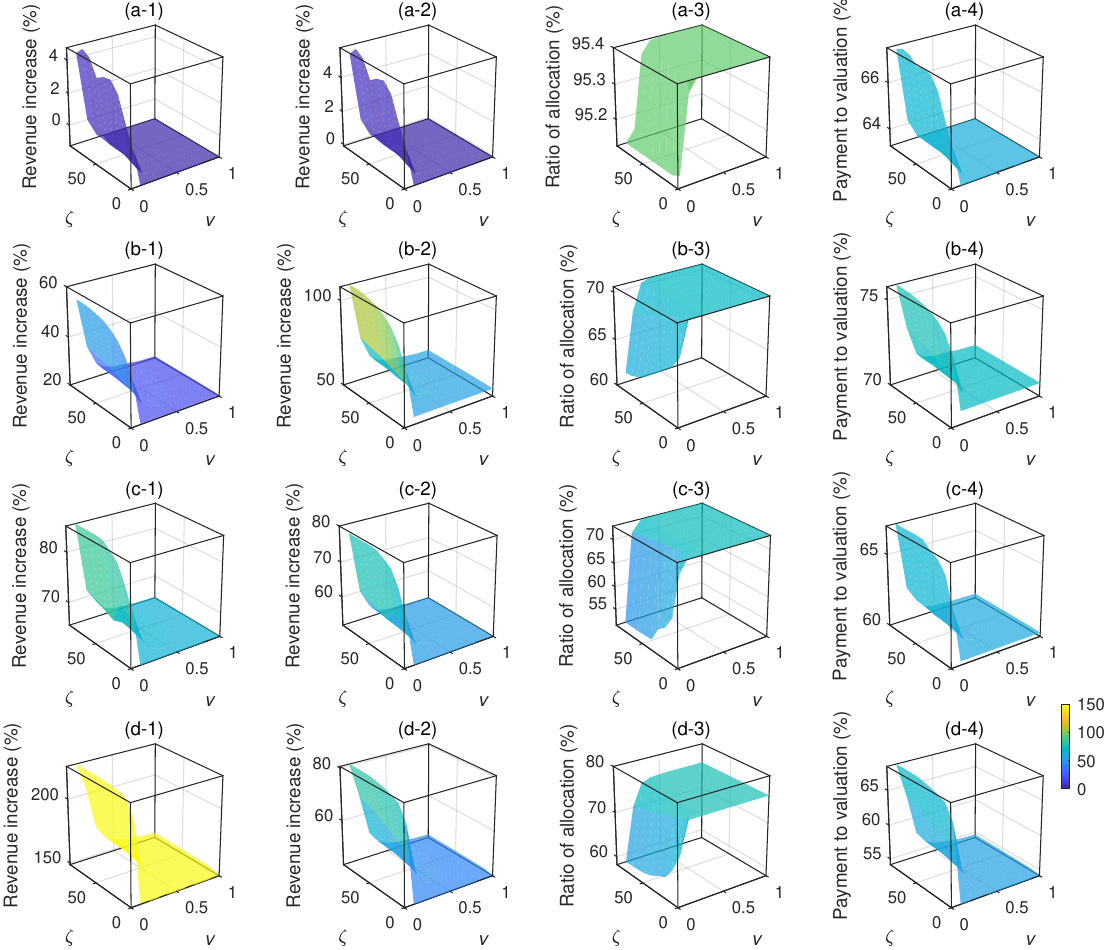}
\caption{Overall results of the model performance on ad slots in: (a) subgroup 1 of group 1 ; (b) subgroup 2 of group 1; (c) subgroup 1 of group 2; (d) subgroup 2 of group 2. The sub-plots show: (1) the average revenue increase of the model to the expected RTB; (2) the average revenue increase of the model to the actual RTB; (3) the average ratio of selling impressions in advance made by the model; (4) the average ratio of payment to valuation made by the model.}
\label{fig:overall_results_clustering}
\end{figure*}

\begin{figure*}[t]
\centering
\includegraphics[width=0.7\linewidth]{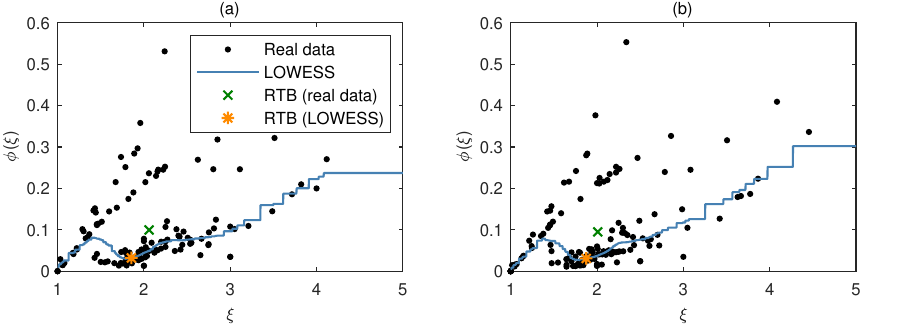}
\caption{Examples of $\phi(\xi)$ estimation for subgroup 2 of: (a) ad slot 14; (b) ad slot 19.}
\label{fig:example_subgroup_payment_difference}
\end{figure*}

\subsection{Advertiser segmentation}
\label{sec:advertiser_segmentation}

Advertisers who bid for the same ad slot can have different valuations on the impressions~\citep{Abraham_2013,Sayedi_2018}. As discussed in Section~\ref{sec:rtb}, truth-telling is a weakly dominant strategy in RTB. Thus we can segment advertisers with different valuations based on their bids. For each ad slot, we use the K-Means clustering to divide the RTB campaigns into two subgroups: subgroup 1 represents impressions bided by advertisers who have high valuations while subgroup 2 represents impressions with low-value advertisers. Table~\ref{tab:stats_groups_clustering} presents a brief summary of the major statistics of subgroups of all 26 ad slots. Similar to Table~\ref{tab:stats_groups}, data expresses similar patterns in both training and test sets so that we can use the training set to develop prediction and pricing models for the future impressions in the test set. In group 1 (where slots have with a high competition level), subgroup 1 has the average 14 advertisers bidding per RTB auction, much higher than that of subgroup 2. In group 2 (where slots have a low competition level), subgroup 1\rq{}s competition level is 4 or 5, almost double subgroup 2. The winning bid and the payment price of subgroup 1 are all higher than those of subgroup 2 because they are positively correlated with the competition level. One interesting finding is that, in group 1, the ratio of payment price to winning bid in subgroup 1 is close to subgroup 2, all around 60\%. This is because the winning advertisers offer high bids in most of RTB campaigns of ad slots 23-25 (whose ratios are around 30\%) though the rest three slots in group 1 have the ratios around 95\%.

Fig.~\ref{fig:example_clustering} presents examples of optimal pricing, allocation and selling of guaranteed contracts for the subgroups of ad slots 3 and 15. We use ad slot 3 to replace ad slot 24 to represent group 1 because impressions from the latter are all suggested to be sold via guaranteed contracts by our model so it is impossible for visualising the optimal allocation. Although small price fluctuations can be seen in some cases, there is a general trend of price increase over time in both subgroups in both markets. Impressions with different values are sold at totally different prices. It is worth emphasising that, in group 1, more high-valued impressions are encouraged to be sold in advance and the total revenue is mainly contributed by guaranteed contracts. This is not only because of the high competition level but also because advertisers with high valuations have great potential to accept a high payment price. Therefore, as shown in Fig.~\ref{fig:example_clustering} (a-1), price increases quickly and then be censored by the upper bound.

Fig.~\ref{fig:overall_results_clustering} summarises the overall results of the model performance on the subgroups of all 26 ad slots. Fig.~\ref{fig:overall_results_clustering} (a-3) shows the average $\gamma$ of subgroup 1 in group 1 is about 95\%, further confirming the replacement of ad slot 24 in Fig.~\ref{fig:example_clustering}. The performance metrics of the model are close for the rest of cases: (i) low-valued impressions in group 1; (ii) high-valued impressions in group 2; (iii) low-valued impressions in group 2. The extreme high revenue increase in Fig.~\ref{fig:overall_results_clustering} (d-1) is due to several outliers and we further explain the reason in Fig.~\ref{fig:example_subgroup_payment_difference}. We use the LOWESS method to fit the $\xi-\phi$ relationship and use the fitted $\phi$ to compute the expected RTB revenue for benchmark as well as for optimisation. As shown in Fig.~\ref{fig:example_subgroup_payment_difference}, the estimated $\phi$ is smaller than the average payment in the real data. Therefore, the expected RTB revenue is smaller than the actual RTB revenue, which then gives extremely high revenue increases on several ad slots.

\section{Conclusion}
\label{sec:conclusion}

In this paper, we study a novel approach of selling display ad impressions via both guaranteed contracts and RTB. We take into account the buying behaviour of advertisers, i.e., risk aversion and stochastic arrivals, and employ dynamic programming to find an optimal pricing and allocation strategy that maximises the publisher's expected total revenue. Our optimal solution is derived based on the solution frame to the Knapsack problem, and is relatively scalable and efficient. We further validate the results using a commercial RTB dataset. The experimental results show that, selling via both guaranteed contracts and RTB can significantly improve the publisher\rq{}s total revenue. Furthermore, for impressions from ad slots with a high competition level, a large percentage of future impressions should be sold in advance via guaranteed contract, and for impressions from ad slots with a low competition level, the revenue is largely collected by RTB. However, their revenue increases are more significant. This is mainly due to the fact that advertisers pay much less than their valuations in RTB which gives more margins for PG to increase the revenue. In addition, the experimental results show that our model is robust under supply or demand uncertainty and when advertisers have different valuations on the impression.

This study has some limitations, which can be addressed as future research directions. First, due to the complex nature of combining both channels, we focus on a simplified model setup in the paper, while keeping the major features of guaranteed contracts and RTB. Our model setup can be relaxed to the case when there are multiple separate groups of impressions with different delivery time periods. For instance, the impressions from one ad slot on a specific day can be treated as a separate group, then for each group, our model can be applied. Therefore, an advertiser can buy impressions from the same ad slot with different delivery time separately. Second, we discuss a simple standardised guaranteed contract, in which the advertiser has no further control of advertising delivery in the delivery period. This setting is consistent with the exiting literature as reviewed in Section~\ref{sec:related_work}. It is possible that an advertiser would like to gain further control of advertising, e.g., deciding when to advertise in the future delivery period. Flexible guaranteed contracts like ad options can be the suitable mechanisms to use~\citep{Chen_2015_1,Chen_2015_2,Chen_2019_1}. However, in these studies, the authors only investigate the contract pricing but not both allocation and pricing. Also, the ad option pricing models discussed in those studies are not optimal. Developing a new revenue maximisation model for optimal pricing and allocation of ad options can be an interesting future topic. Third, when guaranteed contracts are sold, how to prioritise the contracts in the delivery period is not discussed in our paper. This is a different problem that is out of the scope of our study because our model focuses on the sales framework (i.e., the optimal allocation of the future inventories into two channels and the corresponding optimal guaranteed contract prices). In the delivery period, the publisher can give equal priority to the guaranteed contracts or prioritise some contracts based on the pre-sold guaranteed contract price or other metrics~\citep{Feldman_2009,Ghosh_2009}. Therefore, another interesting future direction is developing an unified model which integrates guaranteed contract selling and premium impressions delivery operation. Finally, our model simplifies the advertiser\rq{}s behaviour and decision by using a ratio that represents the proportion of those who want to buy an impression, and we do not explicitly model the advertisers' strategic behaviour of deliberation over PG and RTB. Future research can model the strategic behaviour of advertisers using a utility function~\citep{Su_2007,Aviv_2008}, and study how such a strategic behaviour impacts the publisher\rq{}s optimal pricing and allocation decisions.


\bibliography{pg}

\begin{thebibliography}{63}
\providecommand{\natexlab}[1]{#1}
\providecommand{\url}[1]{\texttt{#1}}
\expandafter\ifx\csname urlstyle\endcsname\relax
  \providecommand{\doi}[1]{doi: #1}\else
  \providecommand{\doi}{doi: \begingroup \urlstyle{rm}\Url}\fi

\bibitem[Abraham et~al.(2013)Abraham, Athey, Babaioff, and Grubb]{Abraham_2013}
I.~Abraham, S.~Athey, M.~Babaioff, and M.~Grubb.
\newblock {Peaches, lemons, and cookies: designing auction markets with
  dispersed information}.
\newblock \emph{Proceedings of the 14th ACM Conference on Electronic Commerce},
  pages 8--9, 2013.

\bibitem[Anjos et~al.(2004)Anjos, Cheng, and Currie]{Anjos_2004}
M.~Anjos, R.~Cheng, and C.~Currie.
\newblock Maximizing revenue in the airline industry under one-way pricing.
\newblock \emph{Journal of the Operational Research Society}, 55\penalty0
  (5):\penalty0 535--541, 2004.

\bibitem[Anjos et~al.(2005)Anjos, Cheng, and Currie]{Anjos_2005}
M.~Anjos, R.~Cheng, and C.~Currie.
\newblock Optimal pricing policies for perishable products.
\newblock \emph{European Journal of Operational Research}, 166\penalty0
  (1):\penalty0 246--254, 2005.

\bibitem[Aviv and Pazgal(2008)]{Aviv_2008}
Y.~Aviv and A.~Pazgal.
\newblock {Optimal pricing of seasonal products in the presence of
  forward-looking consumers}.
\newblock \emph{Manufacturing \& Service Operations Management}, 10\penalty0
  (3):\penalty0 337--562, 2008.

\bibitem[Babaioff et~al.(2009)Babaioff, Hartline, and Kleinberg]{Babaioff_2009}
M.~Babaioff, J.~Hartline, and R.~Kleinberg.
\newblock {Selling ad campaigns: online algorithms with cancellations}.
\newblock \emph{Proceedings of the 10th ACM Conference on Electronic Commerce},
  pages 61--70, 2009.

\bibitem[Balseiro et~al.(2014)Balseiro, Feldman, Mirrokni, and
  Muthukrishnan]{Balseiro_2014}
S.~Balseiro, J.~Feldman, V.~Mirrokni, and M.~Muthukrishnan.
\newblock {Yield optimization of display advertising with ad exchange}.
\newblock \emph{Management Science}, 60\penalty0 (12):\penalty0 2886--2907,
  2014.

\bibitem[{Ben-Zwi} et~al.(2015){Ben-Zwi}, Henzinger, and
  Loitzenbauer]{Ben-Zwi_2015}
O.~{Ben-Zwi}, M.~Henzinger, and V.~Loitzenbauer.
\newblock {Ad exchange: envy-free auctions with mediators}.
\newblock \emph{Proceedings of the 11th Conference Web and Internet Economics},
  pages 104--117, 2015.

\bibitem[Bharadwaj et~al.(2010)Bharadwaj, Ma, Schwarz, Shanmugasundaram, Vee,
  Xie, and Yang]{Bharadwaj_2010}
V.~Bharadwaj, W.~Ma, M.~Schwarz, J.~Shanmugasundaram, E.~Vee, J.~Xie, and
  J.~Yang.
\newblock {Pricing guaranteed contracts in online display advertising}.
\newblock \emph{Proceedings of the 19th ACM International Conference on
  Information and Knowledge Management}, pages 399--408, 2010.

\bibitem[Bharadwaj et~al.(2012)Bharadwaj, Chen, Ma, Nagarajan, Tomlin,
  Vassilvitskii, Vee, and Yang]{Bharadwaj_2012}
V.~Bharadwaj, P.~Chen, W.~Ma, C.~Nagarajan, J.~Tomlin, S.~Vassilvitskii,
  E.~Vee, and J.~Yang.
\newblock {Shale: an efficient algorithm for allocation of guaranteed display
  advertising}.
\newblock \emph{Proceedings of the 18th ACM SIGKDD Conference on Knowledge
  Discovery and Data Mining}, pages 2278--2284, 2012.

\bibitem[Bishop(2006)]{Bishop_2006}
C.~Bishop.
\newblock \emph{Pattern Recognition and Machine Learning}.
\newblock Springer, 2006.

\bibitem[Bitran and Caldentey(2003)]{Bitran_2003}
G.~Bitran and R.~Caldentey.
\newblock {An overview of pricing models for revenue management}.
\newblock \emph{Manufacturing \& Service Operations Management}, 5\penalty0
  (3):\penalty0 179--267, 2003.

\bibitem[Boyle(1986)]{Boyle_1986}
P.~Boyle.
\newblock Option valuation using a three-jump process.
\newblock \emph{International Options Journal}, 3:\penalty0 7--12, 1986.

\bibitem[Caldentey and Vulcano(2007)]{Caldentey_2007}
R.~Caldentey and G.~Vulcano.
\newblock {Online auction and list price revenue management}.
\newblock \emph{Management Science}, 53\penalty0 (5):\penalty0 795--813, 2007.

\bibitem[Chen(2016)]{Chen_2016}
B.~Chen.
\newblock {Risk-aware dynamic reserve prices of programmatic guarantee in
  display advertising}.
\newblock \emph{Proceedings of the 16th IEEE International Conference on Data
  Mining Workshops}, pages 511--518, 2016.

\bibitem[Chen and Kankanhalli(2019)]{Chen_2019_1}
B.~Chen and M.~Kankanhalli.
\newblock {Pricing average price advertising options when underlying spot
  market prices are discontinuous}.
\newblock \emph{IEEE Transactions on Knowledge and Data Engineering},
  31\penalty0 (9):\penalty0 1765--1778, 2019.

\bibitem[Chen and Wang(2015)]{Chen_2015_2}
B.~Chen and J.~Wang.
\newblock {A lattice framework for pricing display advertisement options with
  the stochastic volatility underlying model}.
\newblock \emph{Electronic Commerce Research and Applications}, 14:\penalty0
  465--479, 2015.

\bibitem[Chen et~al.(2014)Chen, Yuan, and Wang]{Chen_2014_2}
B.~Chen, S.~Yuan, and J.~Wang.
\newblock {A dynamic pricing model for unifying programmatic guarantee and
  real-time bidding in display advertising}.
\newblock \emph{Proceedings of the 8th International Workshop on Data Mining
  for Online Advertising}, pages 1--9, 2014.

\bibitem[Chen et~al.(2015)Chen, Wang, Cox, and Kankanhalli]{Chen_2015_1}
B.~Chen, J.~Wang, I.~Cox, and M.~Kankanhalli.
\newblock {Multi-keyword multi-click advertisement option contracts for
  sponsored search}.
\newblock \emph{ACM Transactions on Intelligent Systems and Technology},
  7\penalty0 (5), 2015.

\bibitem[Chen(2017)]{Chen_2017_HKUST}
Y.~Chen.
\newblock {Optimal dynamic auctions for display advertising}.
\newblock \emph{Operations Research}, 65\penalty0 (4):\penalty0 897--913, 2017.

\bibitem[Cleveland(1979)]{Cleveland_1979}
W.~Cleveland.
\newblock Robust locally weighted regression and smoothing scatterplots.
\newblock \emph{Journal of the American Statistical Association}, 74\penalty0
  (368):\penalty0 829--836, 1979.

\bibitem[Constantin et~al.(2009)Constantin, Feldman, Muthukrishnan, and
  P\'{a}l]{Constantin_2009}
F.~Constantin, J.~Feldman, M.~Muthukrishnan, and M.~P\'{a}l.
\newblock {An online mechanism for ad slot reservations with cancellations}.
\newblock \emph{Proceedings of the 20th annual ACM-SIAM Symposium on Discrete
  Algorithms}, pages 1265--1274, 2009.

\bibitem[Constantinides and Malliaris(2001)]{Constantinides_2001}
G.~Constantinides and A.~Malliaris.
\newblock Options markets.
\newblock In \emph{Options Markets Vol I Equity Options Markets: Foundations
  and Pricing}. Edward Elgar Publishing, 2001.

\bibitem[Cox et~al.(1979)Cox, Ross, and Rubinstein]{Cox_1979}
J.~Cox, S.~Ross, and M.~Rubinstein.
\newblock Option pricing: {a} simplified approach.
\newblock \emph{Journal of Financial Economics}, 7:\penalty0 229--263, 1979.

\bibitem[{DoubleClick}(2005)]{DoubleClick_2005}
{DoubleClick}.
\newblock The decade in online advertising 1994-2004.
\newblock \emph{{White Paper}}, 2005.

\bibitem[Edelman et~al.(2007)Edelman, Ostrovsky, and Schwarz]{Edelman_2007_2}
B.~Edelman, M.~Ostrovsky, and M.~Schwarz.
\newblock {Internet} advertising and the generalized second-price auction:
  {selling} billions of dollars worth of keywords.
\newblock \emph{American Economic Review}, 97\penalty0 (1):\penalty0 242--259,
  2007.

\bibitem[eMarketer(2013)]{eMarketer_2013_RTB}
eMarketer.
\newblock Real-time bidding poised to make up quarter of all display spending.
\newblock \url{http://goo.gl/04Ykx}, 2013.

\bibitem[Feldman et~al.(2009)Feldman, Korula, Mirrokni, Muthukrishnan, and
  P{\'a}l]{Feldman_2009}
J.~Feldman, N.~Korula, V.~Mirrokni, M.~Muthukrishnan, and M.~P{\'a}l.
\newblock {Online ad assignment with free disposal}.
\newblock \emph{Proceedings of the 5th International Workshop on Internet and
  Network Economics}, pages 374--385, 2009.

\bibitem[Feldman et~al.(2010)Feldman, Mirrokni, Muthukrishnan, and
  P\'{a}i]{Feldman_2010}
J.~Feldman, V.~Mirrokni, S.~Muthukrishnan, and M.~P\'{a}i.
\newblock Auctions with intermediaries.
\newblock \emph{Proceedings of the 11th ACM conference on Electronic Commerce},
  pages 23--32, 2010.

\bibitem[Fridgeirsdottir and {Najafi-Asadolahi}(2018)]{Fridgeirsdottir_2018_1}
K.~Fridgeirsdottir and S.~{Najafi-Asadolahi}.
\newblock {Cost-per-impression pricing for display advertising}.
\newblock \emph{Operations Research}, 66\penalty0 (3):\penalty0 597--892, 2018.

\bibitem[Gallego and {van Ryzin}(1994)]{Gallego_1994}
G.~Gallego and G.~{van Ryzin}.
\newblock Optimal dynamic pricing of inventories with stochastic demand over
  finite horizons.
\newblock \emph{Management Science}, 40\penalty0 (8):\penalty0 999--1020, 1994.

\bibitem[Gao et~al.(2016)Gao, Shou, Chen, and Huang]{Gao_2016}
L.~Gao, B.~Shou, Y.-J. Chen, and J.~Huang.
\newblock {Combining spot and futures markets: a hybrid market approach to
  dynamic spectrum access}.
\newblock \emph{Operations Research}, 64\penalty0 (4):\penalty0 794--821, 2016.

\bibitem[Ghosh et~al.(2009)Ghosh, McAfee, Papineni, and
  Vassilvitskii]{Ghosh_2009}
A.~Ghosh, P.~McAfee, K.~Papineni, and S.~Vassilvitskii.
\newblock {Bidding for representative allocations for display advertising}.
\newblock \emph{Proceedings of the 5th International Workshop on Internet and
  Network Economics}, pages 14--18, 2009.

\bibitem[Hojjat et~al.(2014)Hojjat, Turner, Cetintas, and Yang]{Hojjat_2014}
A.~Hojjat, J.~Turner, S.~Cetintas, and J.~Yang.
\newblock Delivering guaranteed display ads under reach and frequency
  requirements.
\newblock \emph{Proceedings of the 28th AAAI Conference on Artificial
  Intelligence}, pages 2278--2284, 2014.

\bibitem[Ilfeld and Winer(2002)]{Ilfeld_2002}
J.~Ilfeld and R.~Winer.
\newblock {Generating website traffic}.
\newblock \emph{Journal of Advertising Research}, 42\penalty0 (5):\penalty0
  49--61, 2002.

\bibitem[Kleinberg and Tardos(2005)]{Kleinberg_2005}
J.~Kleinberg and E.~Tardos.
\newblock \emph{Algorithm Design}.
\newblock Addison Wesley, 2005.

\bibitem[Lahaie and McAfee(2011)]{Lahaie_2011}
S.~Lahaie and P.~McAfee.
\newblock {Efficient ranking in sponsored search}.
\newblock \emph{Proceedings of the 7th International Conference on Internet and
  Network Economics}, pages 254--265, 2011.

\bibitem[Lahaie and Pennock(2007)]{Lahaie_2007}
S.~Lahaie and D.~Pennock.
\newblock {Revenue analysis of a family of ranking rules for keyword auctions}.
\newblock \emph{Proceedings of the 8th ACM conference on Electronic commerce},
  pages 50--60, 2007.

\bibitem[Lee and Leckenby(1999)]{Lee_1999}
S.~Lee and J.~Leckenby.
\newblock {Impact of measurement periods on website rankings and traffic
  estimation: a user-centric approach}.
\newblock \emph{Journal of Current Issues and Research in Advertising},
  21\penalty0 (2):\penalty0 1--10, 1999.

\bibitem[Mansour et~al.(2012)Mansour, Muthukrishnan, and Nisan]{Mansour_2012}
Y.~Mansour, S.~Muthukrishnan, and N.~Nisan.
\newblock {Doubleclick ad exchange auction}, 2012.
\newblock {arXiv:} \url{https://arxiv.org/abs/1204.0535}.

\bibitem[McGill and {van Ryzin}(1999)]{McGill_1999}
J.~McGill and G.~{van Ryzin}.
\newblock {Revenue management: Research overview and prospects}.
\newblock \emph{Transportation Science}, 33\penalty0 (2):\penalty0 233--256,
  1999.

\bibitem[Muthukrishnan(2009)]{Muthukrishnan_2009}
M.~Muthukrishnan.
\newblock {Ad exchanges: research issues}.
\newblock \emph{Proceedings of the 5th International Workshop on Internet and
  Network Economics}, pages 1--12, 2009.

\bibitem[Myerson(1981)]{Myerson_1981}
R.~Myerson.
\newblock Optimal auction design.
\newblock \emph{Mathematics of Operational Research}, 6\penalty0 (1):\penalty0
  58--73, 1981.

\bibitem[{Najafi-Asadolahi} and Fridgeirsdottir(2014)]{Najafi-Asadolahi_2014}
S.~{Najafi-Asadolahi} and K.~Fridgeirsdottir.
\newblock {Cost-per-click pricing for display advertising}.
\newblock \emph{Manufacturing \& Service Operations Management}, 16\penalty0
  (4):\penalty0 482--497, 2014.

\bibitem[Narahari(2014)]{Narahari_2014}
Y.~Narahari.
\newblock \emph{Game Theory and Mechanism Design}.
\newblock World Scientific, 2014.

\bibitem[{OpenX}(2013)]{OpenX_2013}
{OpenX}.
\newblock Programmatic + premium: {c}urrent practices and future trends.
\newblock \emph{White Paper}, 2013.

\bibitem[Ostrovsky and Schwarz(2011)]{Ostrovsky_2011}
M.~Ostrovsky and M.~Schwarz.
\newblock {Reserve prices in Internet advertising auctions: a field
  experiment}.
\newblock \emph{Proceedings of the 12th ACM Conference on Electronic Commerce},
  pages 59--60, 2011.

\bibitem[Parkes(2007)]{Parkes_2007}
D.~Parkes.
\newblock \emph{Algorithmic Game Theory}, chapter Online Mechanisms.
\newblock Cambridge University Press, 2007.

\bibitem[Radovanovic and Heavlin(2012)]{Radovanovic_2012}
A.~Radovanovic and W.~Heavlin.
\newblock {Risk-aware revenue maximization in display advertising}.
\newblock \emph{Proceedings of the 21st International Conference on World Wide
  Web}, pages 91--100, 2012.

\bibitem[Roels and Fridgeirsdottir(2009)]{Roels_2009}
G.~Roels and K.~Fridgeirsdottir.
\newblock Dynamic revenue management for online display advertising.
\newblock \emph{Journal of Revenue \& Pricing Management}, 8\penalty0
  (5):\penalty0 452--466, 2009.

\bibitem[Salomatin et~al.(2012)Salomatin, Liu, and Yang]{Salomatin_2012}
K.~Salomatin, T.~Liu, and Y.~Yang.
\newblock {A unified optimization framework for auction and guaranteed delivery
  in online advertising}.
\newblock \emph{Proceedings of the 21st ACM International Conference on
  Information and Knowledge Management}, pages 2005--2009, 2012.

\bibitem[Sayedi(2018)]{Sayedi_2018}
A.~Sayedi.
\newblock {Real-time bidding in online display advertising}.
\newblock \emph{Marketing Science}, 37\penalty0 (4):\penalty0 553--568, 2018.

\bibitem[Su(2007)]{Su_2007}
X.~Su.
\newblock {Intertemporal pricing with strategic customer behavior}.
\newblock \emph{Management Science}, 53\penalty0 (5):\penalty0 726--741, 2007.

\bibitem[Sun et~al.(2016)Sun, Dawande, Janakiraman, and Mookerjee]{Sun_2016}
Z.~Sun, M.~Dawande, G.~Janakiraman, and V.~Mookerjee.
\newblock {The making of a good impression: information hiding in ad
  exchanges}.
\newblock \emph{MIS Quarterly}, 40\penalty0 (3):\penalty0 717--739, 2016.

\bibitem[Talluri and {van Ryzin}(2005)]{Talluri_2004}
K.~Talluri and G.~{van Ryzin}.
\newblock \emph{The Theory and Practice of Revenue Management}.
\newblock Springer, 2005.

\bibitem[Thompson and {Leyton-Brown}(2013)]{Thompson_2013}
D.~Thompson and K.~{Leyton-Brown}.
\newblock {Revenue optimization in the generalized second-price auction}.
\newblock \emph{Proceedings of the 14th ACM conference on Electronic Commerce},
  pages 837--852, 2013.

\bibitem[Varian(2007)]{Varian_2007}
H.~Varian.
\newblock Position auctions.
\newblock \emph{International Journal of Industrial Organization}, 25\penalty0
  (6):\penalty0 1163--1178, 2007.

\bibitem[Varian(2009)]{Varian_2009}
H.~Varian.
\newblock Online ad auctions.
\newblock \emph{American Economic Review}, 99\penalty0 (2):\penalty0 430--434,
  2009.

\bibitem[Varian and Harris(2014)]{Varian_2014}
H.~Varian and C.~Harris.
\newblock {The VCG auction in theory and practice}.
\newblock \emph{American Economic Review}, 104\penalty0 (5):\penalty0 442--445,
  2014.

\bibitem[Wang and Chen(2012)]{Wang_2012_1}
J.~Wang and B.~Chen.
\newblock {Selling futures online advertising slots via option contracts}.
\newblock \emph{Proceedings of the 21st International World Wide Web
  Conference}, pages 627--628, 2012.

\bibitem[Wilmott(2006)]{Wilmott_2006_1}
P.~Wilmott.
\newblock \emph{Paul Wilmott On Quantitative Finance}.
\newblock John Wiley, 2nd edition, 2006.

\bibitem[Yuan et~al.(2014{\natexlab{a}})Yuan, Wang, Chen, Mason, and
  Seljan]{Yuan_2014}
S.~Yuan, J.~Wang, B.~Chen, P.~Mason, and S.~Seljan.
\newblock {An empirical study of reserve price optimisation in real-time
  bidding}.
\newblock \emph{Proceedings of the 20th ACM SIGKDD Conference on Knowledge
  Discovery and Data Mining}, pages 1897--1906, 2014{\natexlab{a}}.

\bibitem[Yuan et~al.(2014{\natexlab{b}})Yuan, Wang, Li, and Qin]{YYuan_2014}
Y.~Yuan, F.~Wang, J.~Li, and R.~Qin.
\newblock {A survey on real time bidding advertising}.
\newblock \emph{IEEE International Conference on Service Operations and
  Logistics, and Informatics}, pages 418--423, 2014{\natexlab{b}}.

\bibitem[Zhang et~al.(2017)Zhang, Wang, Li, Zhang, Lan, Li, and
  Sun]{Zhang_2017}
J.~Zhang, Z.~Wang, Q.~Li, J.~Zhang, Y.~Lan, Q.~Li, and X.~Sun.
\newblock {Efficient delivery policy to minimize user traffic consumption in
  guaranteed advertising}.
\newblock \emph{Proceedings of the 31st AAAI Conference on Artificial
  Intelligence}, pages 253--258, 2017.

\end{thebibliography}

\newpage
\onecolumn
\section*{Appendix A: Notations}

This appendix is not an essential part of the paper but may help readers reference the key notations used throughout the paper.\\

\begin{table}[h]
\centering
\caption{Summary of the key notations}
\begin{tabular}{l|p{5.5in}}
\toprule
Notation & Description\\
\hline
$t_0, \cdots, t_N, t_{\widetilde{N}}$ & Discrete time points: $[t_0, t_N]$ is the period to sell the guaranteed contracts; $[t_N, t_{\widetilde{N}}]$ is the period that the impressions are created, auctioned off (in RTB) and delivered.\\
$Q$ & Estimated total demand for impressions in $[t_N, t_{\widetilde{N}}]$.\\
$S$ & Estimated total supply of impressions in $[t_N, t_{\widetilde{N}}]$.\\
$R^{PG}$ & Expected revenue from selling guaranteed contracts.\\
$R^{RTB}$ & Expected revenue from selling the remaining impressions in RTB.\\
$p(t_n)$ & Price of a guaranteed contract sold at time $t_n$, $n=0,\cdots,N$.\\
$\gamma$ & Ratio of allocation of impressions into guaranteed contracts. \\
$\omega$ & Probability that the publisher fails to deliver a guaranteed impression in $[t_N, t_{\widetilde{N}}]$.\\
$\varpi$ & Size of penalty so the publisher needs to pay $\varpi p(t_n)$ penalty if he fails to deliver a guaranteed impression which is sold at $p (t_n)$.\\
$\eta (t_n)$ & Total (accumulative) arrived but unfulfilled demand at time $t_n$.\\
$\theta(t_n, p(t_n))$ & Proportion of advertisers who are willing to buy an impression in advance at time $t_n$ and at price $p(t_n)$.\\
$\alpha$ & Price effect in $\theta(t_n, p(t_n))$.\\
$\beta$ & Time effect in $\theta(t_n, p(t_n))$.\\
$\delta(t_n)$ & Risk preference for a buyer at time $t_n$.\\
$\zeta$ & Risk level in $\delta(t_n)$.\\
$v$ & Time effect in $\delta(t_n)$.\\
$\xi(t_N)$ & Per-auction competition level of RTB in $[t_N, t_{\widetilde{N}}]$.\\
$\phi(\xi(t_n))$ & Expected payment from an impression in RTB for the given $\xi(t_n)$.\\
$\psi(\xi(t_n))$ & Expected risk of an impression in RTB for the given $\xi(t_n)$.\\
$\chi(t_n, \xi(t_n))$ & Risk-aware upper bound of the guaranteed contract price at time $t_n$.\\
$\Phi(t_n)$ & Censored upper bound of the guaranteed contract price at time $t_n$.\\
$\pi$ & Expected maximum value on an impression.\\
$\lambda$ & Intensity of the Poisson process describing the demand arrivals.\\
$f(t_n)$ & Expected number of arrivals in $[t_{n-1}, t_{n}]$.\\
$\mathcal{O}$ & Execution time required by an algorithm.\\
$\mathbb{E}[\cdot]$ & Expectation. \\
$\mathrm{var}[\cdot]$ & Variance.\\
$g(\cdot)$ & Density function of advertiser's bid.\\
$F(\cdot)$ & Cumulative distribution function of advertiser's bid.\\
$u_n$ & Sales upper bound at time $t_n$ in Algorithm~\ref{algo:solution}.\\
$l_n$ & Sales lower bound at time $t_n$ in Algorithm~\ref{algo:solution}.\\
$\mathcal{Y}_n$ & Set of sales at time $t_n$ in Algorithm~\ref{algo:solution}.\\
$\widetilde{\mathbf{H}}_n(y)$ & $R^{PG}$ when selling $y$ guaranteed contracts up to time $t_n$ in Algorithm~\ref{algo:calc_hn}, $y \in \mathcal{Y}_n$.\\
$\mathbf{H}_n(y)$ & Optimal $R^{PG}$ in Algorithms~\ref{algo:solution}-\ref{algo:calc_hn}, $y \in \mathcal{Y}_n$.\\
$R(y)$ & Total revenue of selling $y$ guaranteed contracts up to time $t_n$ in Algorithm~\ref{algo:solution}, $y \in \mathcal{Y}_n$.\\
\bottomrule
\end{tabular}
\end{table}

\newpage
\section*{Appendix B: Examples of optimal pricing and allocation}

\begin{figure}[h]
\centering
\includegraphics[width=1\linewidth]{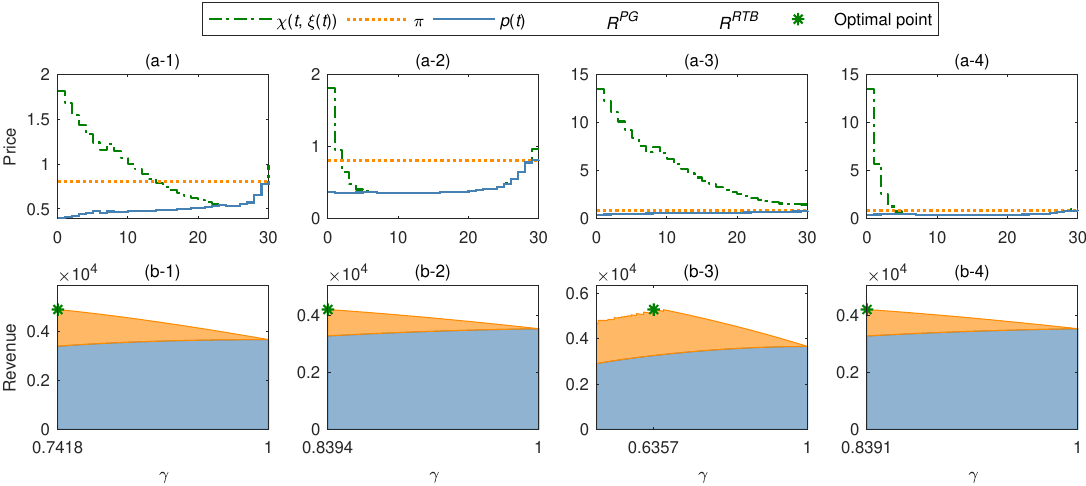}
\caption{Examples for ad slot 7: (a) optimal pricing; and (b) allocation of guaranteed contracts. Parameters are set differently in the subplots: (1) $\zeta = 10, v = 0.1$; (2) $\zeta = 10, v=0.9$; (3) $\zeta = 90, v = 0.1$; (4) $\zeta = 90, v = 0.9$.}
\label{fig:example_pricing_and_allocation_01}
\vspace*{15pt}
\includegraphics[width=1\linewidth]{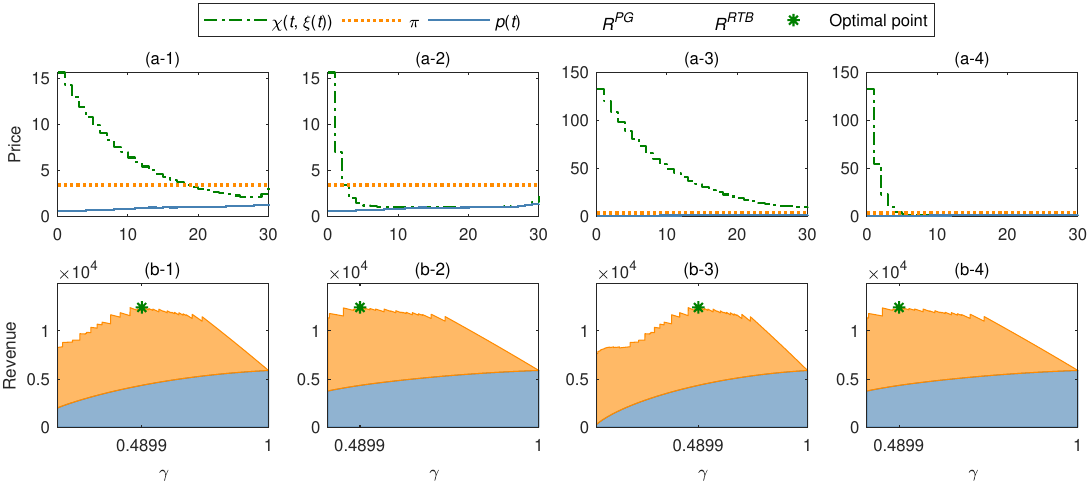}
\caption{Examples for ad slot 11: (a) optimal pricing; and (b) allocation of guaranteed contracts. Parameters are set differently in the subplots: (1) $\zeta = 10, v = 0.1$; (2) $\zeta = 10, v=0.9$; (3) $\zeta = 90, v = 0.1$; (4) $\zeta = 90, v = 0.9$.}
\label{fig:example_pricing_and_allocation_02}
\end{figure}

Here we provide two additional examples to justify our analysis in Section~\ref{sec:optimal_pricing_and_allocation}. Fig.~\ref{fig:example_pricing_and_allocation_01} shows examples from ad slot 7 where the effects of $\zeta$ and $v$ on optimal allocation are significant while Fig.~\ref{fig:example_pricing_and_allocation_02} shows examples from ad slot 11 where their effects are not significant.\\

\end{document}